\documentclass{JHEP3}
\pdfoutput=1
\usepackage[utf8]{inputenc}
\usepackage{bbm,slashed}
\usepackage{graphicx,epsfig}
\usepackage{amsmath,amsfonts,amssymb}  
\newcommand{\psib}{\bar{\psi}} 
    
\newcommand{\Dbar}{\bar{D}}

\newcommand{\thetab}{\bar{\theta}}
\newcommand{\pa}{\partial}  
\newcommand{\pat}{\partial_\theta} 
\newcommand{\patb}{\partial_{\bar\theta}}
\newcommand{\eps}{\epsilon}
\newcommand{\epsb}{\bar\epsilon}

\newcommand{\ha}{\frac{1}{2}}

\DeclareMathOperator{\STr}{STr}

\DeclareMathAlphabet{\boldmathe}{T1}{cmr}{bx}{it}

\newcommand{\cQ}{{\mathcal Q}}

\newcommand{\mtxt}[1]{\quad\hbox{{#1}}\quad}

\def\suint{\text{d}\theta \text{d}\thetab}
\graphicspath{{plots/}}

\title{Convergence of Derivative Expansion in Supersymmetric Functional RG Flows} 
\author{Marianne Heilmann,  Tobias Hellwig, Benjamin Knorr, Marcus Ansorg and Andreas
  Wipf\\
  Theoretisch-Physikalisches Institut, Friedrich-Schiller-Universit{\"a}t
Jena,
Max-Wien-Platz 1, D-07743 Jena, Germany\\
   E-mail: \email{marianne.heilmann@uni-jena.de}, \email{tobias.hellwig@uni-jena.de}, 
   \email{benjamin.knorr@uni-jena.de}, \email{marcus.ansorg@uni-jena.de}, \email{wipf@tpi.uni-jena.de}}

\abstract{We confirm the convergence of the derivative expansion in two supersymmetric models
via the functional renormalization group method. Using pseudo-spectral 
methods, high-accuracy results for the lowest energies in supersymmetric quantum mechanics 
and  a detailed description of the supersymmetric analogue of the Wilson-Fisher
fixed point of the three-dimensional Wess-Zumino model are obtained. The superscaling 
relation proposed earlier, relating the relevant critical exponent to the 
anomalous dimension, is shown to be valid to all orders in the supercovariant
derivative expansion and for all $d \ge 2$.} 
  
 \preprint{}
  
 \keywords{Renormalization Group, Effective Action, Superspace, Supersymmetric Effective
 Theories, Nonperturbative Effects, Quantum Mechanical Tunneling, Pseudo-spectral methods}

\begin{document}

\section{Introduction}
Supersymmetry (SUSY) is an essential ingredient in most theories
beyond the Standard Model of particle physics. Under
certain natural conditions it is the unique extension of the
Poincar\'e symmetry. Although many predictions of supersymmetry can be 
explored by perturbative calculations there remain
interesting phenomena which fall into the non-perturbative regime. 
Examples are collective condensation phenomena, topological effects 
and supersymmetry breaking. Thus, powerful methods are needed
to investigate and calculate such non-perturbative effects in 
order to understand the underlying physics and to obtain
quantitative predictions in the strong-coupling regime.

A well-established method to study non-perturbative effects
in supersymmetric theories is based on a discretization of 
spacetime and the corresponding supersymmetric lattice
models, see e.g.  
\cite{ Feo:2004kx,  Giedt:2006pd, Catterall:2007kn, Bergner:2007pu,
Takimi:2007nn, Wozar:2011gu, Baumgartner:2012np, Bergner:2013jia}. Yet, as
supersymmetry involves spacetime translations and rotations, lattice approaches usually go along with a complete (for models
with extended supersymmetry partial)
loss of supersymmetry. 

In this work, we choose an alternative approach based on 
the functional renormalization group (FRG). The FRG is a powerful 
non-perturbative tool to study continuum physics and was successfully 
applied to a plethora of physical systems which include
fermionic systems, gauge theories or gravity, see e.g. \cite{Wetterich:1992yh,Tetradis:1993ts,
Reuter:1993kw,Gies:2010st,Herbst:2014zea,Christiansen:2014raa}. Only lately, the
FRG has been applied to SUSY models -  mainly due to the difficulty of
constructing a sensible regularization that preserves SUSY. Here, we will extend previous 
results \cite{Synatschke:2008pv,Synatschke:2009nm,Synatschke:2010ub} by examining whether the derivative 
expansions for supersymmetric quantum mechanics or supersymmetric Yukawa theories
are convergent, and further investigate SUSY breaking in these models.
To that aim we admit higher order supercovariant derivative terms
in the effective actions of supersymmetric Wess-Zumino-models in various 
dimensions.

In the early days of the FRG, ordinary quantum mechanics was already 
utilized as toy model to test the power of the techniques beyond perturbation 
theory. In (supersymmetric) quantum mechanics, flow equations are easily 
obtained and solutions to truncated flow equations can be compared with 
exact results. The focus in these studies were the energies of
the low lying states \cite{Horikoshi:1998sw,Kapoyannis:2000sp,
Zappala:2001nv,Gies:2006wv,Wipf:2013vp}. It turned out that systems with single-well potentials 
can be successfully described all the way from weak to strong couplings, whereas double-wells
are more challenging, partially due to exponentially suppressed instanton effects.

The cited works satisfactorily calculated the energy spectrum of non-supersymmetric
systems. In non-supersymmetric systems there is no need to relate the
regulators of bosons and fermions. However, in a supersymmetric systems
they must be related by supersymmetry in order to disentangle the effects 
of spontaneous and explicit supersymmetry breaking. Thus in the present
work we use manifestly supersymmetric regulators such that for supersymmetric
initial conditions the scale dependent effective action is supersymmetric at 
all scales. This way we considerably improve upon previous results 
obtained in \cite{Synatschke:2008pv}. 
Other works using a supersymmetric regularization include 
\cite{Synatschke:2009da,Heilmann:2012yf,Litim:2011bf}.

Regarding the models investigated in the present work, we shall see that the
supercovariant derivative expansion converges nicely.
 To calculate the flow of the superpotential and wave functions
renormalizations to fourth order in this expansion with 
high precision we employ the powerful spectral method \cite{Boyd:ChebyFourier}.
This method has been  very successfully applied to problems in hydrodynamics,
quantum chemistry and in particular gravity \cite{Ansorg:2003br}.
To solve the system of coupled nonlinear partial differential equations
with spectral methods we use  the Chebyshev polynomials as basis in the domain 
where the effective potential is flat or concave and rational Chebyshev functions 
in the domains where the effective potential is convex. For the RG-evolution,
another Chebyshev spectralization was used. This way we are
able to construct \emph{global solutions} 
to the truncated flow equations with unmatched numerical accuracy.

The paper is organized as follows: In section \ref{sec:susyQM} we review
the relevant features of SUSY quantum mechanics. Sections \ref{sec:flowderivation} and 
\ref{sec:Eff} contain the derivation of the flow equations in
next-to-next-to-leading order (NNLO) as well as numerical results for the first 
excited energies in systems with unbroken supersymmetry. Spontaneous
supersymmetry breaking is analyzed in next-to-leading order (NLO) in section
\ref{sec:SUSYBreaking}. In section \ref{sec:Wess-Zumino} we proceed with the derivative expansion of the three-dimensional Wess-Zumino models. 
The truncated flow equations are analysed in section \ref{sec:Wess-Zumino}, which
contains an analysis of the physics of the Wilson-Fisher fixed point 
and the superscaling relation previously discovered in \cite{Gies:2009az}. The appendices 
contain further details on the derivation of the flow equations and on the spectral 
method used in the present work.
\newpage
\section{Supersymmetric quantum mechanics}
\label{sec:susyQM}
In order to derive the flow equations for supersymmetric quantum mechanics, 
we employ the superfield formalism \cite{Salam:1974yz}. For more details
we refer the reader to \cite{Synatschke:2008pv}. The  Euclidean superfield, 
expanded in terms of the anticommuting Grassmann variables $\theta$ 
and $\bar{\theta}$, reads%
\begin{align}
\Phi(\tau, \theta, \bar{\theta})=\phi(\tau) + \thetab\psi(\tau) +\psib
\theta(\tau) +\thetab\theta F(\tau).\label{susy1}
\end{align}
Both the superfield and the Grassmann variables $\theta, \bar{\theta}$ have
mass dimension $-1/2$.
Next, we introduce the superpotential and expand it in powers of the
Grassmann variables,
\begin{align}
W(\Phi)=W(\phi)+\big(\thetab\psi+\psib\theta\big)W'(\phi)+
\thetab\theta\big(F W'(\phi)-W''(\phi)\psib\psi\big).\label{susy4}
\end{align}
The one-dimensional equivalent of the Super-Poincar{\'e} algebra 
contains only translations of Euclidean time and is generated by one pair 
of conserved nilpotent fermionic supercharges $\cQ=i\patb+\theta\pa_\tau$ {and} 
$\bar \cQ=i\pat+\thetab\pa_\tau$. The anticommutator of them is the super-Hamiltonian,
\begin{align}
\{\cQ, \bar{\cQ}\}=2H, \;\quad [H,\cQ]=[H,\bar{\cQ}]=0.\label{susy41}
\end{align}
Supersymmetry variations are generated by
$\delta_\eps=\epsb \cQ-\eps\bar \cQ$.
We may easily read off the following transformation rules of the component fields
\begin{align}
\delta\phi=i\epsb\psi-i\psib\eps\,, \quad
\delta\psi=(\dot\phi-iF)\eps\,,\quad
\delta\psib=\epsb(\dot\phi+iF),\quad
\delta F=-\epsb\dot\psi-\dot{\psib}\eps\, ,
\label{susy11}
\end{align}
by acting with the  supersymmetry variations on the superfield:
\begin{align}
\delta_\eps\Phi
=\epsb\big(i\psi+i\theta F+\theta\dot\phi-\thetab\theta\dot\psi\big)
-\big(i\psib+i\thetab F
-\thetab\dot\phi+\thetab\theta\dot{\psib}\big)\eps.\label{susy9} 
\end{align}
Here and in the following, a dot denotes differentiation w.r.t. $\tau$.
In order to obtain a supersymmetric action, we further need the 
\emph{supercovariant derivatives}
$D=i\patb-\theta\pa_\tau$ and $\bar D=i\pat-\thetab\pa_\tau$. They
fulfill almost identical anticommutation relations as the supercharges,
\begin{align}
\{D,D\}=\{\bar D,\bar D\}=0\mtxt{and}\{D,\bar D\}=-2H,\label{susy13}
\end{align}
and anticommute with the  supercharges.
With these definitions, one can write down
the supersymmetric Euclidean \emph{off-shell action} within the superfield formalism:
\begin{align}
   	S[\phi,F,{\psib},\psi]&=
    \int \text{d}\tau \suint
  	\left[-\frac12\Phi
	K\Phi+i W(\Phi)\right]\notag\\
& =
  \int \text{d}\tau
  	\left[\frac{1}{2}\dot\phi^2-i\psib\dot\psi
  	+iFW'(\phi)-i\psib W''(\phi)\psi+\frac12 F^2\right],\label{susy19}
  	\end{align}
where we introduced the  kinetic operator 
\begin{equation}
K=\ha(\Dbar D - D \Dbar).\label{kinetic}
\end{equation}
A prime always denotes the derivative with respect to the scalar field $\phi$.
Eliminating the auxiliary field $F$ by its equation of motion, $F=-i W'$, 
we obtain the \emph{on-shell action}
\begin{align}
	S_{\rm on}[\phi,\psi,\psib]
	=\int \text{d}\tau\left[\frac{1}{2}\dot\phi^2-i\psib\dot\psi 
	+\frac12 W'{}^2(\phi)-W''(\phi)\psib\psi
	\right].\label{susy21}
\end{align}
From \eqref{susy21} we read off the  bosonic potential 
$V(\phi)=\frac12W'{}^2(\phi)$ and the
Yukawa term $W''\psib\psi$. If supersymmetry 
is unbroken, the ground state energy vanishes.

Let us assume the superpotential $W(\phi)$ to be a polynomial in the scalar field. 
Then, the global properties of the superpotential $W(\phi)\sim \phi^n$ for 
large $\phi$  determine whether spontaneous breaking of supersymmetry 
occurs or not. If $n$ is even, supersymmetry will be intact on all scales. 
This is realized e.g. for quartic classical superpotentials
\begin{align}
W(\phi)=e\phi+\frac m2\phi^2+\frac g3\phi^3+\frac a4\phi^4\,,\label{susy23}
\end{align}
which we will consider in section \ref{subsec:numerics}.
$W(\phi)$ represents 
the microscopic superpotential, i.e. the initial potential of our quantum 
system before fluctuations are taken into account.
If $n$ is odd, the effective potential exhibits a ground state with positive
energy and supersymmetry is spontaneously broken, even if  we may start with a 
microscopic potential with vanishing ground state energy. This applies
e.g. to  cubic classical superpotentials of the form
\begin{equation}
W(\phi)=e\phi+\frac{g}{3}\phi^3,\quad\quad\; e<0,\;g>0,
\label{potBreaking}
\end{equation}
which will be discussed in detail in section \ref{sec:SUSYBreaking}.
\subsection{Flow equation in superspace}
\label{sec:flowderivation}
In order to analyze supersymmetric quantum 
mechanical systems, we resort to Wilsonian renormalization group techniques. 
Specifically, we adopt the framework of the FRG, formulated in terms 
of a flow equation for the effective average action $\Gamma_k$. 
It is based on the infinitesimal integrating-out of degrees of freedom 
with momenta larger than some infrared momentum scale $k^2$. Thus, 
at a certain scale $k$ all quantum fluctuations with momenta $\vert p\vert>k$ are 
taken into account. Hence, $\Gamma_k$ interpolates between the microscopic action $S$ 
in the ultraviolet and the full quantum effective action $\Gamma_{k\rightarrow0}=\Gamma$ 
in the infrared (IR).
It obeys the exact functional differential equation \cite{Wetterich:1992yh}
\begin{equation}
 \partial_k\Gamma_k=
 \frac12 \STr\left\{\left[\Gamma_k^{(2)}+ R_k\right]^{-1}\partial_k  R_k\right\}.
\label{eq:wetterich}
\end{equation}
This flow equation reads in superspace
\begin{equation}
\partial_k \Gamma_k =\frac{1}{2}\int \,\text{d}z \, \text{d}z^{\prime}\partial_k R_k(z,z')G_k(z',z), \quad\quad G_k=(\Gamma^{(2)}_k+R_k)^{-1}, 
\label{flowsusy}
\end{equation}
where  $z= (\tau,\theta, \bar{\theta})$ denotes  the coordinates in superspace. 
Therein the second functional derivative with respect to the superfield 
$\Gamma^{(2)}_k$ is given by
\begin{equation}
(\Gamma^{(2)}_k)(z,z^{\prime})=\frac{\overrightarrow{\delta}}{\delta\Phi(z)}\Gamma_k\frac{\overleftarrow{\delta}}{\delta\Phi(z^{\prime})}.
\label{G2}
\end{equation}
Note that the supertrace in \eqref{eq:wetterich} as well as the right- and left-derivatives in \eqref{G2} take care of the minus signs for anticommuting variables.

\subsubsection{Supercovariant derivative expansion in NNLO}
In this work, we employ the expansion of $\Gamma_k$ in powers of the
supercovariant derivatives $D$ and $\bar{D}$ with mass-dimension $1/2$. 
Unfortunately, a systematic and consistent expansion scheme of $\Gamma_k$ 
does not guarantee convergence. One goal of the present work is
to demonstrate the convergence of the supercovariant derivative expansion
at NNLO to numerically known values 
of observables.
We will derive the flow equation in the off-shell formulation with a manifestly 
supersymmetric regulator such that in each order of the 
supercovariant derivative expansion the flow preserves supersymmetry. \\
To this order, the most general ansatz for the scale-dependent effective action
reads
\begin{align}
  \Gamma_k[\Phi]&= \int \text{d}z \left[i\, W_k(\Phi)-\frac12 Z_k(\Phi)
    K Z_k(\Phi)+\frac{i}{4}Y_{1,k}(\Phi)K^2\Phi+\frac{i}{4}Y_{2,k}(\Phi)(K\Phi)(K\Phi)\right]\, ,
    \label{flow0}
\end{align}
with the scale and field dependent functions $W_k,Z_k,Y_{1,k}$ and $Y_{2,k}$ and 
the kinetic operator $K$ introduced in (\ref{kinetic}).
A contribution to $\Gamma_k$ where the derivatives act on three superfields,
$Y_{3,k} (\bar{D}\Phi) (D\Phi) (K\Phi)$, is already included in our truncation,
since
\begin{equation}
\int \text{d}z\, A(\Phi) K^2\Phi=\int \text{d}z\, \left[A'(\Phi) (K\Phi)(K\Phi)+ A''(\Phi)
(\bar{D}\Phi) (D\Phi) (K\Phi)\right].
\label{relation}
\end{equation}
Terms with $D, \bar{D}$ acting on four superfields do not exist since $(\bar{D}\Phi D\Phi)^2=0$.
In component fields the action (\ref{flow0}) takes the 
form\footnote{For convenience, we often omit the index $k$ and the explicit dependence on
the superfield $\Phi$ of the scalar
functions $W_k, Z_k,Y_{1,k}$ and $Y_{2,k}$ from now on.}
\begin{align}
 \Gamma_k[\Phi]   
    &= \int \text{d}\tau
  \left[\frac{1}{2}Z'^2\dot{\phi}^2-i Z^{\prime 2} \bar{\psi}\dot{\psi}-\frac{i}{2}
  \big(Y_1'+Y_2\big)\dot{\bar{\psi}}\dot{\psi}
  -i\big(W'' +Z'Z''\dot{\phi}-\frac{1}{2}Y_1''\ddot{\phi}-\frac{1}{4}Y_1'''\dot{\phi}^2\big)\bar{\psi}\psi\right.\notag\\
  &\quad+\left(i W'-Z'Z'' \bar{\psi}\psi-\frac{i}{2}\big(Y_1'+Y_2\big)\ddot{\phi}-\frac{i}{4}Y_1''\dot{\phi}^2+\frac{1}{2}Y_2'\big(\bar{\psi}\dot{\psi}-\dot{\bar{\psi}}\psi\big)\right)F\notag\\
  &\quad\left.+\left(\frac{1}{2}Z^{\prime 2}- \frac{i}{4}Y_2''\bar{\psi}\psi\right)F^2
  +\frac{i}{4}Y_2'F^3
  \right]\,,\label{flow1}
\end{align}
where the terms are ordered according to increasing powers of the auxiliary field $F$.

\subsubsection{Introducing the regulator functional}

The flow of $\Gamma_k$ is regularized by adding a suitable 
regulator functional $\Delta S_k$ to the action, in such a way that $R_k=\Delta S_k^{(2)}$. 
Given a supersymmetric truncation $\Gamma_k$ and a supersymmetric initial 
condition, we only need a supersymmetric regulator in order to 
construct a manifestly supersymmetric flow. Following 
\cite{Synatschke:2008pv, Synatschke:2009nm, Synatschke:2010ub}, the most 
general off-shell supersymmetric cutoff action quadratic in the superfields 
can be written as
\begin{align}
\Delta S_k =\frac12  \int \text{d}z\, \Phi R_k(D,\Dbar)
\Phi\,.\label{reg1} 
\end{align}
As $D$ and $\bar{D}$ satisfy the anticommutation relation \eqref{susy13},
it can be written as
\begin{align}
\Delta S_k =
\frac12  \int \text{d}z\, \Phi\left[ir_1(-\partial_\tau^2,k)-Z'{}^2(\bar{\Phi})\,r_2(-\partial_\tau^2,k)K\right]\Phi, \label{reg3}
\end{align} 
where $Z'$ is evaluated at the background field $\bar{\Phi}=\bar{\phi}$. 
Thus, in momentum space $R_k$ is given by\footnote{We abbreviate $\delta(\theta,\theta'):=\delta(\bar{\theta}-\bar{\theta}^{\prime})\delta(\theta-\theta^{\prime})$.} 
\begin{equation}
R_k(q,q',\theta, \theta') =
	\left[ir_1(q^2,k)-Z'{}^2(\bar{\phi})\,r_2(q^2,k)
	K(q,\theta)\right]\delta(q,q^{\prime})\delta(\theta,\theta')\,.
\label{RK}
\end{equation} 
The regulator function $r_1$ with mass dimension $1$ acts like an additional
momentum-dependent mass and ensures a gap $\sim k$ for the IR modes. Note that we do
not spectrally adjust this regulator function by multiplying it with
the wave function renormalization as has been done in \cite{Synatschke:2008pv}. The
latter approach would actually slow down the flow of the higher
order operators $Z, Y_{1}, Y_{2}$. The dimensionless regulator function $r_2$
can be viewed as a deformation of the momentum dependence of the kinetic 
term. The term $q^2r_2(q^2/k^2)$ represents the supersymmetric 
analogue of the corresponding regulator function $r_k(q^2/k^2)$ in scalar 
field theory \cite{Tetradis:1993ts}. Here, a spectral adjustment via the 
inclusion of the wave function renormalization $Z'(\bar{\Phi})$ is helpful 
in order to provide a simple form for the flow of $\Gamma_k$ \cite{Litim:2001up}.
We did check the influence of the spectral adjustment on the flow of $\Gamma_k$ 
carefully.
 
\subsubsection{Flow equation} 
We begin with the calculation of the second functional derivative of $\Gamma_k$  
as defined in \eqref{G2} in order to derive its flow according to
\eqref{flowsusy}. We find\footnote{The functional derivative w.r.t. 
the superfields is defined such that $\int dz \frac{\delta \Phi(z)}{\delta \Phi(z^{\prime})}=\int dz \delta(z,z^{\prime})=1$,  where 
$\delta(z,z^{\prime}):=\delta(\tau,\tau^{\prime})\delta(\theta,\theta')$.} 
\begin{align}
&\left(\Gamma_k^{(2)}+R_k\right)(z,z')= \left[i(W''+ r_1)-Z''(KZ)-Z'KZ'
-Z'(\bar{\Phi})^2r_2K+\frac{i}{4}\Big\{Y_1''(K^2\Phi)\right.\notag\\
&\hskip5mm\left.+Y_1'K^2+K^2Y_1'+Y_2''(K\Phi)^2+2Y_2'(K\Phi)K+2KY_2'(K\Phi)+2KY_2K\Big\}\right]\delta(z,z').
\label{Gamma2}
\end{align} 
The scale dependent functions $W,Z,Y_1, Y_2$  are functions of the superfield 
$\Phi(z)$ whereas the scale dependent $Z'(\bar\Phi)$  has the background field as argument.
Here, a bracket implies that the kinetic operator $K$ only acts within the bracket.
If there is no bracket, then it acts on everything on its right hand side. 

We may derive the flow of the scalar functions  $W'(\phi), Z'(\phi)$ and
$Y_2'(\phi)$ via a projection onto the coefficients of $F, F^2$ 
and $F^3$ in \eqref{flow1} with fermionic fields and time derivatives set to zero.
Thus, it suffices to consider constant component fields in \eqref{Gamma2} and 
set $\bar{\psi}=\psi=0$ afterwards. Switching to momentum
space, the inverse propagator \eqref{Gamma2} takes the form
\begin{align}
&\left(\Gamma_k^{(2)}+R_k\right)(q, q',\theta, \theta')
=\left[\Big(i\big(W''+r_1\big)+Z'Z''F +\frac{i}{2}\big(Y_1'+Y_2\big) q^2+\frac{i}{4}Y_2''F^2\Big)\delta(\theta, \theta')\right.\notag\\
&+\Big(iW'''F+\big(Z'Z'''+Z''^2\big)F^2+ B q^2+\frac{i}{4}Y_2'''F^3+i\big(\frac{1}{2}Y_1''+Y_2'\big)F q^2\Big)\bar{\theta}\theta\bar{\theta}'\theta'\label{Gamma22} \\
&\left.+\big(B+\frac{3}{2}i F Y_2'\big)+\big(B q+i F Y_2'q\big)(\bar{\theta}'\theta-\bar{\theta}\theta')+\big(Z'Z''F+\frac{i}{2}Y_2''F^2\big)(\bar{\theta}\theta+\bar{\theta}'\theta')\right]\delta(q,q'),
\notag
\end{align}
where the background field enters via $B=Z'^2+r_2Z'^2(\bar{\phi})$.
The Greens function in superspace, $G_k=(\Gamma_k^{(2)}+R_k)^{-1}$, is determined by
\begin{equation}
\int  \frac{\text{d}q'}{2\pi}\, \text{d}\theta'\, \text{d}\bar{\theta}'\,G_k^{-1}(q,q', \theta,
\theta')\,G_k(q',q'',\theta',\theta'')=\delta(q, q'')\,\delta(\theta, \theta'').
\label{defGreen}
\end{equation}
To continue, we make the  general ansatz
\begin{equation}
G_k(q,q',\theta,\theta')=\left(a+b\,\bar{\theta}\theta+c\,\bar{\theta}'\theta'+d\,\bar{\theta}\theta'+e\,\bar{\theta}'\theta+f\, \bar{\theta}\theta\bar{\theta}'\theta'\right)\delta(q,q') \, ,
\label{ansatz}
\end{equation}
with arbitrary coefficients depending on the scalar functions
in \eqref{Gamma22} and the momentum $q$. By solving \eqref{defGreen} for the 
coefficients in $G_k$, the projected flow equation reads:
\begin{align}
&\left.\partial_k \Gamma_k\right|_{\dot{\phi}=\dot{F}=\psi=\bar{\psi}=0}= \int \text{d}\tau 
\left(i \partial_k W'F+\frac{1}{2}\partial_kZ'{}^2 F^2+\frac{i}{4}\partial_kY_2' F^3\right)\notag\\
&=\frac{1}{2}\int \frac{\text{d}q}{2\pi}\,\frac{\text{d}q'}{2\pi}\,\text{d}\theta \,\text{d}\bar{\theta}\,\text{d}\theta'\,\text{d}\bar{\theta}'(\partial_kR_k)(q',q,\theta',\theta)G_k(q,q',\theta,\theta')\notag\\
&=\frac{1}{2}\int \text{d}\tau\,\frac{\text{d}q}{2\pi}\left[i(\partial_kr_1)(b+c+d+e)+\partial_k\big(r_2Z'^2(\bar{\phi})\big)(f+a q^2-eq +dq)\right].
\label{fluss1}
\end{align} 
The solutions for the coefficients are presented in appendix A.
By extracting the coefficients of $F,F^2$ and $F^3$ on the right hand side
we obtain the flow equations for $W',Z'$ and $Y_2'$. Since the flow of $Y_1$ is missing, we further project 
the flow equation onto the  coefficient of $\dot{F}\dot{\phi}$. This 
way, one obtains the flow of $Y_1'(\phi)+Y_2(\phi)$  and thus a closed 
system of four coupled PDE's. Appendix B contains the detailed derivation 
of this flow which requires the
inverse propagator for momentum-dependent fields $\phi,F$.
To simplify the obtained flow equations, we define the new functions
\begin{equation}
Y:=Y_2'\quad\hbox{and}\quad
X:=Y_1'+Y_2.
\label{newY}
\end{equation}
It remains to solve the flow equations for the scale dependent functions $W',Z',X$ and $Y$.

\subsection{Effective potential and first excited energy}
\label{sec:Eff}
The low lying energies can be extracted from the bosonic on-shell effective potential
$V_\mathrm{eff}=V_{k=0}$.
In order to compute $V_{k}$, we set  the fermionic fields to zero in the 
truncated effective average action  \eqref{flow1}.
\subsubsection{On-shell effective potential}
At a given scale, the auxiliary field $F$ fulfills the equation of motion
\begin{equation}
F=-\frac{2i}{3Y}\left(\sqrt{Z'^4+\frac{3}{4}\big(4W'-2X\ddot{\phi}
-(X'-Y)\dot{\phi}^2\big)Y}-Z'^2\right) \, ,
\label{EOMF}
\end{equation}
and hence becomes dynamical, in contrast to the situation in the NLO 
approximation. Next, we eliminate the auxiliary field in the bosonic action
by its equation of motion. To calculate the effective potential $V_\mathrm{eff}$, 
it is sufficient to consider $\Gamma_k[\phi]$ for constant $\phi$ in which case
\begin{equation}
V_k(\phi)=\frac{2}{27 Y^2}\big(\sqrt{3 W'Y+Z'^4}-Z'^2\big)
\big(6W'Y+Z'^4-Z'^2\sqrt{3 W'Y+Z'^4}\big)\,.
\label{potential}
\end{equation}   
We determine the energy of the first  excited state $E_1$ from 
the propagator $G_k$ at vanishing $k$, where the regulator $R_k$ vanishes. 
Supersymmetry is unbroken if the potential $V_k$ in (\ref{potential}) vanishes 
at its minimum $\phi_{\min}$, which is the case
if $W_k'(\phi_{\min})=0$. Actually, in the strong coupling regime
there exists a second solution for which $[4 W'(Y+Z'^4)](\phi_{\min})=0$.
However, we believe this solution to be  unphysical, see section
\ref{subsec:numerics}.\\ For a constant $\phi_\mathrm{min}$, the auxiliary field $F$ in \eqref{EOMF}
vanishes if $W_k'(\phi_{\min})=0$. Thus, we determine 
the excited energies $E_1$ by considering the propagator \eqref{ansatz} 
for constant fields $\phi$ and $W'=F=0$. After an integration over the Grassmann 
variables, we obtain
\begin{equation}
\left.G_k(q,q', \theta, \theta')\right|_{\bar{\theta}\theta\,\bar{\theta}'\theta'}=\frac{Z'^2 q^2}{Z'^4 q^2+(W''+1/2 X q^2)^2}\delta(q-q').
\label{propsuper}
\end{equation}
The square of the excited energy $E_1^2$ is then given by the pole of the 
propagator at the minimum of the effective potential:
\begin{equation}
\lim_{k\rightarrow0}\left(Z'^4q_0^2+(W''+\frac{1}{2}X q_0^2)^2\right)\Big|_{\phi_{\min}}=0\quad  \mbox{with}\quad q_0^2=(iE_1)^2.
\label{pole}
\end{equation}
This equation possesses the two solutions
\begin{equation}
E_1^2=\lim_{k\rightarrow0}\frac{2}{X^2}\left(Z'^4+X W''\pm Z'^2\sqrt{Z'^4+2X W''}\right)
\Big|_{\phi_{\min}},\label{zerosprop}
\end{equation}
where the solution with the negative sign is the correct one,
since it reduces to the known limiting value $E_1=\vert W''(\phi_\mathrm{min})\vert$ 
in the LPA approximation with $Z'=1$ and $X=0$. The other solution with
positive sign diverges in this limit.\\
Note that if supersymmetry is spontaneously broken, $W_k'(\phi_\mathrm{min})\neq 0$ 
and the corresponding auxiliary field $F$ does not vanish. Then the first excited 
energy $E_1$ is extracted from the pole of the general propagator \eqref{ansatz}, 
i.e. of 
\begin{equation}
\lim_{k\rightarrow 0}\left.G_k(q,q', \theta,
\theta')\right|_{\bar{\theta}\theta\,\bar{\theta}'\theta'}
\label{propallgemein}
\end{equation}
at the (constant) minimum $\phi_\mathrm{min}$ of the potential,
where  $F$ has to be replaced by its equation 
of motion \eqref{EOMF}.

\subsubsection{Numerical results}
\label{subsec:numerics}

The flow equation of the superpotential in NNLO in the derivative expansion is
given in its full form by
\begin{align}
\partial_k W_k'(\phi)\!&=\frac{1}{2}\int\limits_{-\infty}^{\infty} \frac{\text{d}
q}{2\pi}\left[\partial_k r_1\frac{2   \left(Z'Z'' \left(A^2-B^2 q^2
\right)-B A A'\right)}{\left(B^2 q^2+A^2\right)^2}\right.\notag\\
&\left.\phantom{\ldots\ldots\ldots..}+\partial_k \big(r_2 Z'{}^2(\bar\phi)\big)
\frac{\left(A' \left(A^2-B^2 q^2\right)+4 B q^2 A Z'Z''
\right)}{\left(B^2 q^2+A^2\right)^2}\right],
\label{flowwithp}
\end{align}
where $A=W''+r_1+\frac{q^2}{2}X$ and as earlier $B=Z'^2+r_2 Z'{}^2(\bar\phi)$.
Now, we specify the regulator functions  and choose $r_2(q^2,k)=0$ and
the Callan-Symanzik regulator $r_1(q^2,k)=k$. Then there is no
dependence on the background field and the flow equation of 
$W^{\prime}(\phi)$ simplifies to
\begin{equation}
\partial_kW'(\phi)=\frac{Z'}{4\mathcal{W}''^2}\frac{\mathcal{W}''^2 \left(X'Z'+4
X Z''\right)-\mathcal{W}''' Z' \left( 3\mathcal{W}''X+Z'^4\right)}{\left(2 \mathcal{W}'' 
   X+Z'^4\right)^{3/2}},\quad \mathcal{W}''=W''+k,
\end{equation}
where the momentum integration has been performed.
Note that the right hand side of the flow equations only
depends via $W''$ and $W'''$ on the superpotential.
The corresponding microscopic action in the UV is given by \eqref{susy21}
and we focus on quartic superpotentials of the form \eqref{susy23}. Thus, the
initial conditions for the flow at $k=\Lambda$ read
\begin{equation}
W^{\prime}_{\Lambda}(\phi)=e
+m\phi+g\phi^2+a\phi^3,\quad Z_{\Lambda}'(\phi)=1,\quad Y_{\Lambda}(\phi)=X_{\Lambda}(\phi)=0.
\label{UVinitial}
\end{equation}
In supersymmetric quantum mechanics the fluctuations in the UV are suppressed
and the flow freezes out for $k \rightarrow \Lambda \rightarrow \infty$.
Hence the initial conditions are stable for large UV-cutoffs.
Indeed, plugging \eqref{UVinitial} into the flow equations yields
 \begin{equation}
 \partial_kW'\big\vert_{\Lambda}=O\left(\Lambda^{-2}\right),\;\;
 \partial_k Z'\big\vert_{\Lambda}=O\left(\Lambda^{-4}\right),\;\;
 \partial_kX\big\vert_{\Lambda}=O\left(\Lambda^{-5}\right),\;\;
\partial_kY\big\vert_{\Lambda}=O\left(\Lambda^{-6}\right)\,.
 \label{UVlimit}
 \end{equation}
For $W^{\prime}_{\Lambda}$ in (\ref{UVinitial}) supersymmetry remains 
unbroken at all scales. Note that the initial superpotential $W_{\Lambda}$ 
is non-convex if $g^2>3 m a$. Besides, we may shift the field $\phi
\rightarrow \phi-g/(3a)$ such that the quadratic term of $W'_\Lambda$
vanishes. 

We solved the set of the four coupled partial differential equations for $W',Z',X,Y$ numerically with spectral methods (see Appendix \ref{sec:appendix3}). Besides, we 
repeated the calculations with the implicit Runge-Kutta method 
of \sffamily NDSolve \rmfamily of MATHEMATICA 9.
Here, we have chosen $\phi \in (-100,100)$ and kept the  four functions at 
their classical values at the boundary for all scales as the 
flows vanish for $|\phi|\rightarrow \infty$. With both methods we obtained the same results to three or four significant
digits. \\ Table \ref{tab:energygaps} displays the energy gap $E_1(g)$ for $e=a=m=1$ and various 
values of the coupling $g$.
We also listed the resulting energies  obtained by solving the PDE's in LPA 
(includes $W_k(\phi)$) as well as NLO  (includes $W_k(\phi)$ and $Z_k(\phi)$).

\begin{figure*}[ht]      
\centering{
\includegraphics[width=0.485\textwidth]{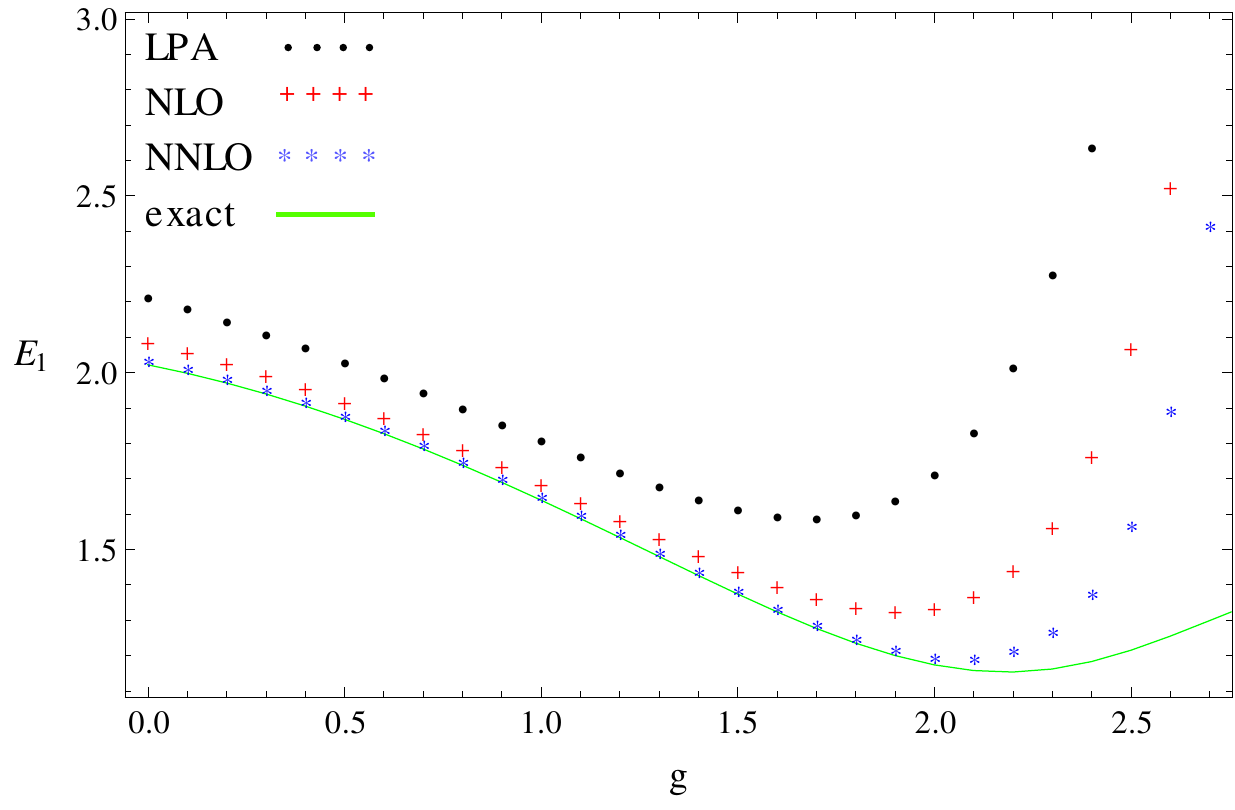}\hfill 
\includegraphics[width=0.5\textwidth]{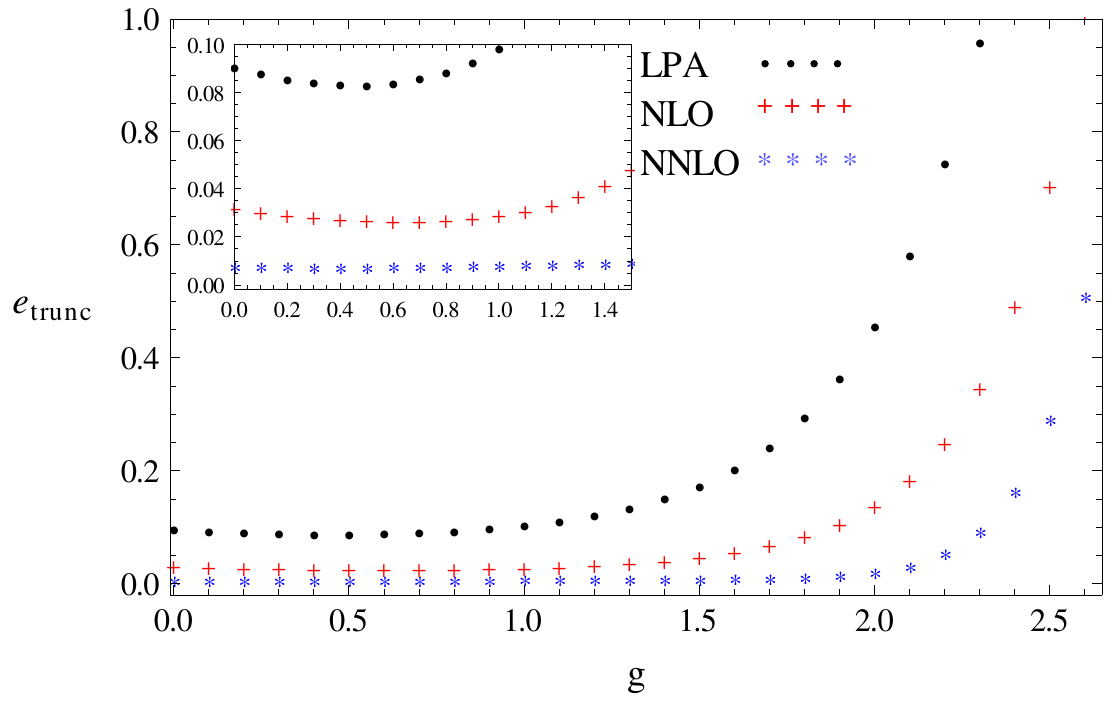}} 
\caption{\label{fig:E1susy}Energy gap $E_1(g)$ and relative error
$e_{\mathrm{trunc}}$ for classical superpotentials of the form
$W'_{\Lambda}(\phi)=1+\phi+g\phi^2+\phi^3$ and various $g$. 
For convex initial potentials ($g<\sqrt{3}$) we achieve a nice convergence 
as well as a relative error of $1\%$ in NNLO.  For couplings  larger than
$g\approx 2$, where the classical potential becomes non-convex, we observe significant deviations 
from the exact results.}
\end{figure*}

Fig.\ref{fig:E1susy} shows the  first excited energies $E_1$ (left figure) and
the relative deviation from the exact values
$e_{\mathrm{trunc}}=(E_1-E_{1}^\mathrm{ex})/E_{1}^\mathrm{ex}$ (right figure) as
a function of the coupling $g$. Obviously, an inclusion of terms of fourth-order in the derivative expansion  improves the results for the energy gap considerably. We obtain a 
relative error of $<1\%$ for  couplings $g< \sqrt{3}$. 
For couplings $\sqrt{3} < g < 2.3$ the relative deviations from the exact
results lie within a  $10\%$ error.

\TABLE{
\begin{tabular}[t]{p{1.5cm}p{1.5cm}p{1.5cm}p{1.5cm}c}
\hline
$g$ & $E_1^\mathrm{LPA}$&$E_1^\mathrm{NLO}$& $E_1^\mathrm{NNLO}$& $E_1^\mathrm{ex}$\\ \hline
0.0& 2.202&2.086&2.038&2.022\\
0.2&2.136&2.028&1.986&1.970\\
0.4&2.061&1.957&1.920&1.905\\
0.6&1.978&1.876&1.842&1.827\\
0.8&1.889&1.784&1.752&1.738\\
1.0&1.797&1.687&1.653&1.639\\
1.2&1.709&1.584&1.547&1.534\\
1.4&1.632&1.486&1.440&1.426\\
1.6&1.583&1.398&1.337&1.323\\
1.8&1.590&1.339&1.250&1.235\\
2.0&1.702&1.337&1.199&1.173\\ 
2.2&2.005&1.442&1.216&1.153\\
2.4&2.627&1.764&1.378&1.183\\ 
2.6&3.661&2.525&1.895&1.254\\ 
2.8&4.988&3.961&3.195&1.343\\ \hline
\end{tabular}
\caption{Energy $E_1^\mathrm{NNLO}$ of the first excited state, calculated according to \eqref{zerosprop} for $r_1=k$, $e=m=a=1$ and various $g$.  For comparison, also the results $E_1^\mathrm{LPA}$ obtained in LPA, $E_1^\mathrm{NLO}$ derived in NLO  as well as  the exact values $E_1^\mathrm{ex}$  from numerically diagonalizing the Hamiltonian are given. Here, $E_1^\mathrm{LPA}, E_1^\mathrm{NLO}$ and $E_1^\mathrm{NNLO}$ were derived by solving the respective partial differential equations numerically.
  \label{tab:energygaps}} 
} 

\begin{figure*}[ht]
\centering{ 
\includegraphics[width=0.5\textwidth]{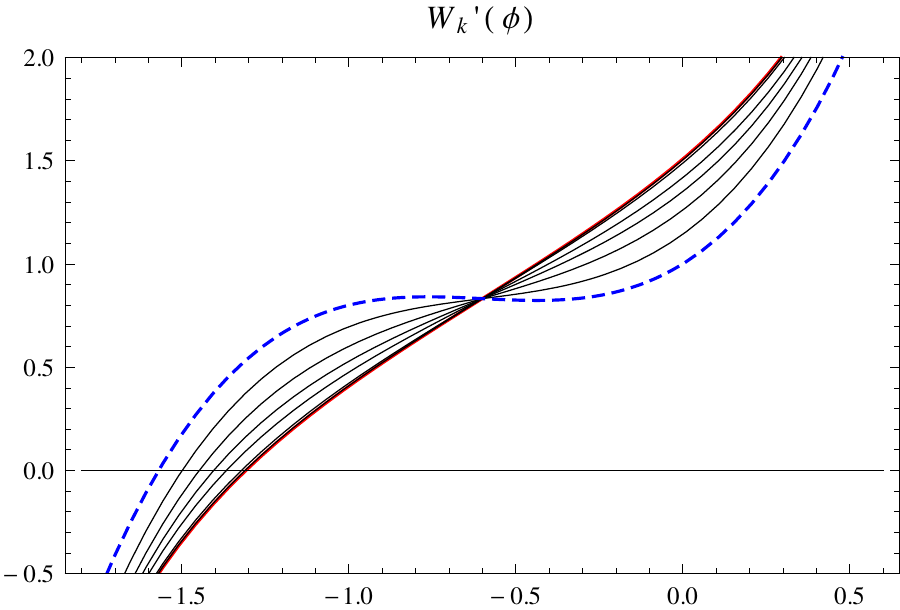}\hfill
\includegraphics[width=0.5\textwidth]{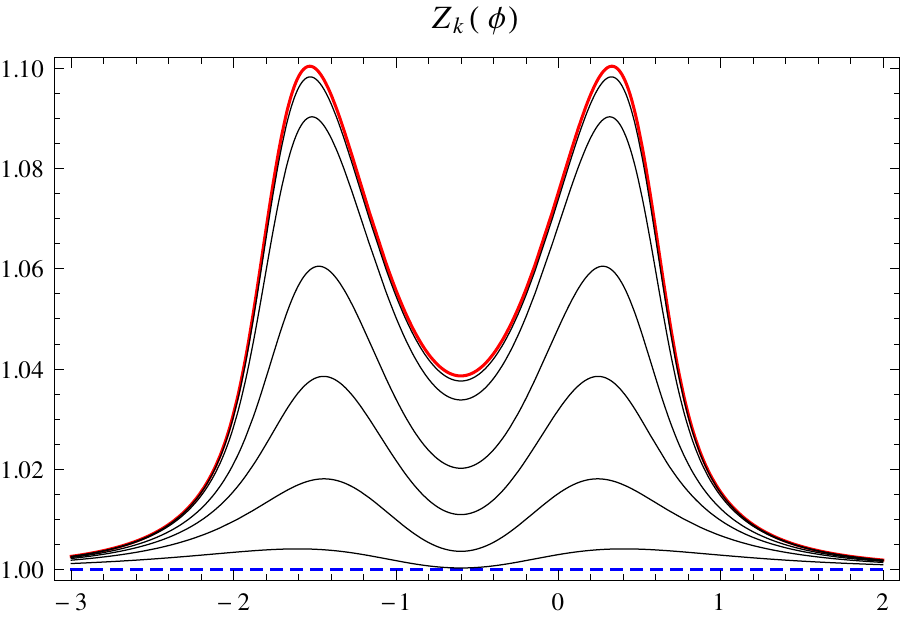}}
\caption{\label{fig:Flows1} The superpotential $W_k'(\phi)$ and the wave function renormalization $Z_k'(\phi)$ for different scales $k$. Starting in the $UV$ at $k=\Lambda=10^4$ (blue, dashed line) the flow evolves to the IR at $k=0$ (red solid line). The intermediate scales are $k=5,2,1,0.5,0.1,0.02$.} 
\end{figure*}

\begin{figure*}[ht]
\centering{
\includegraphics[width=0.5\textwidth]{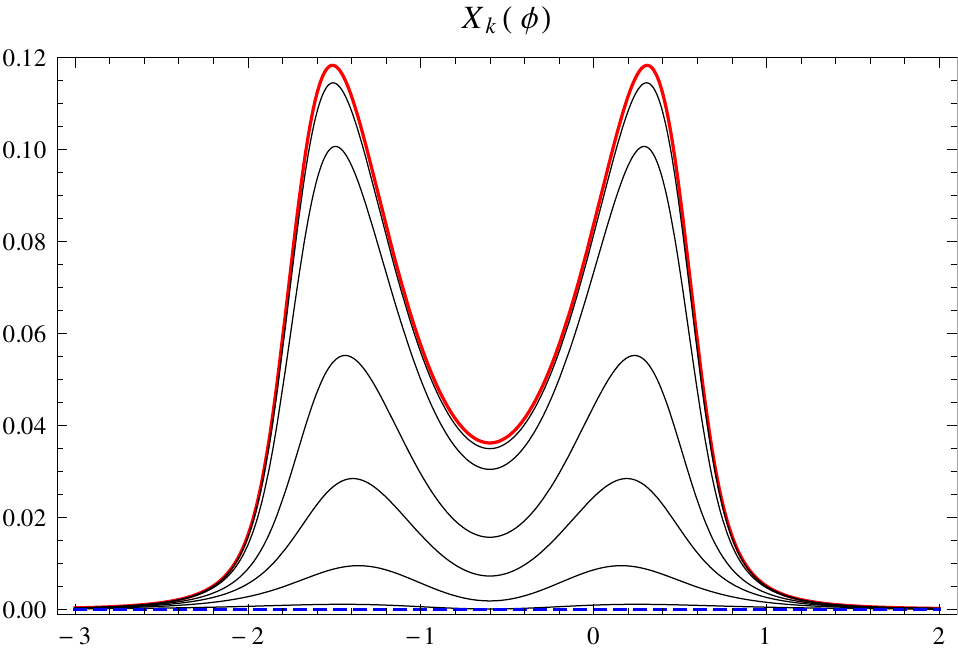}\hfill
\includegraphics[width=0.5\textwidth]{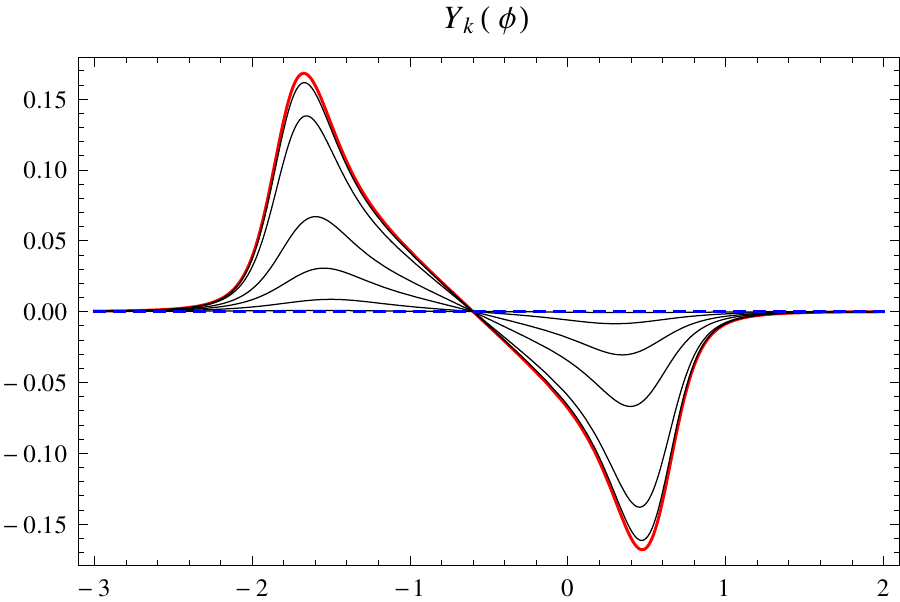}}
\caption{\label{fig:Flows2} The flow of the fourth-order couplings $X_k(\phi),Y_k(\phi)$  for different scales $k$. Starting in the $UV$ at $k=\Lambda=10^4$ (blue, dashed line) the flow evolves to the IR at $k=0$ (red solid line). The intermediate scales are $k=5,2,1,0.5,0.1,0.02$} 
\end{figure*}

Note that for couplings larger than $g\approx 2$ the error 
increases exponentially and the supercovariant derivative approximation
breaks down. The breakdown of the NNLO approximation for couplings $g\gtrsim 2$
is also indicated by the structure of the effective average  potential. In this regime,
$V_k(\phi)$ becomes complex for all scales smaller than a $k_0>0$
for field values close to the local minimum of $W'_{\Lambda}$. This is due to the expression $\sqrt{3W'Y+Z'^4}$ appearing in \eqref{potential} which becomes complex near
 the local minimum of $W'_{\Lambda}$ for  non-convex
initial potentials, owing to an increasingly negative $Y$; see also figures
\ref{fig:Flows1}, \ref{fig:Flows2}. 
Another sign of the breakdown is given by the appearance
of a further mass at $g \approx 1.7$, splitting in two masses for even larger
couplings $g$. This is due to the formation of one/two further minima of the effective
potential, where $\left.4W'Y+Z'^4\right|_{\phi_{\min}}=0$ holds. Here, the
fourth order correction $Y$ is of the same
order as the leading and next-to leading order terms $W'$ and $Z'$ indicating
the invalidity of the truncation. The corresponding masses 
become parametrically large. These large masses in the strong coupling regime 
are probably an artifact of the regularization and have no physical 
significance. Similar artifacts have already been encountered in $O(N)$ symmetric 
Wess-Zumino models \cite{Heilmann:2012yf}.\\
We thus observe a very good convergence of the derivative
expansion in case the local barrier of the classical potential is small. 
However, as the non-convexity of $V_{\Lambda}$ increases,
tunneling events are exponentially suppressed and are no longer 
captured by the flow equations in the derivative expansion.
Here, the inclusion of non-local operators should lead to
a better convergence behaviour in the strong coupling regime. 
 
\begin{figure*}[ht] 
\centering{ 
\includegraphics[width=0.5\textwidth]{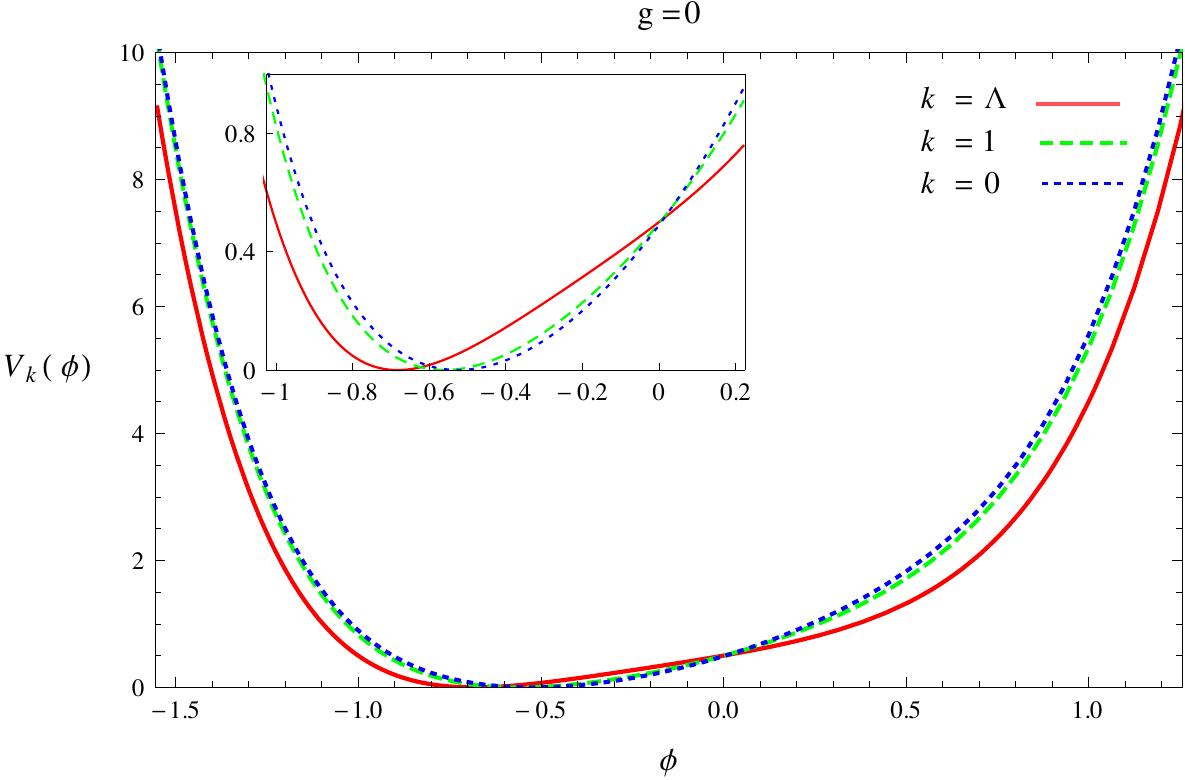}\hfill
\includegraphics[width=0.5\textwidth]{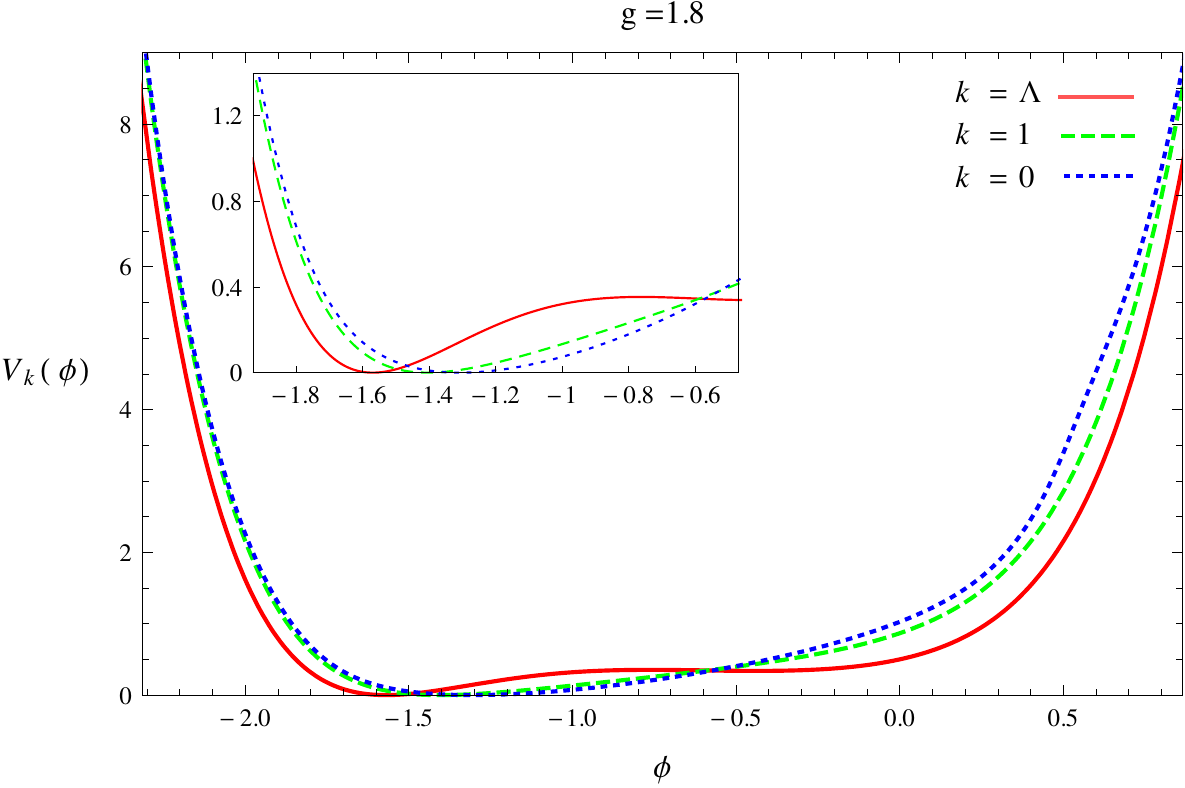}}
\caption[]{\label{fig:Potential} Flow of the effective average potential
$V_k(\phi)$ as obtained by solving the system of PDE's numerically in NNLO 
in the derivative expansion with initial 
conditions \eqref{UVinitial}, where $e=m=a=1$ and $g=0$ (left panel) and 
$g=1.8$ (right panel).} 
\end{figure*}
The flow of the bosonic
potential $V_k(\phi)$ for $g=0$ and $g=1.8$ is depicted in figure \ref{fig:Potential}. Apparently, non-convexities appearing in the classical potential diminish as more and more long-range 
quantum fluctuations are taken into account such that the effective potentials in figure \ref{fig:Potential} become convex.\footnote{However, note that the 
structure of the flow equation \eqref{eq:wetterich} forces rather 
the superpotential than the scalar potential to become convex in the IR. 
Since the scalar potential is a complicated function of $W',Z,Y$, i.e. 
of the form (\ref{potential}), the flow equation does not immediately
imply $V_{k\rightarrow 0}$ to be convex.}
Furthermore, figures \ref{fig:Flows1} and \ref{fig:Flows2} show the 
flow of $W',Z',X$ and $Y$ for $g=1.8$. From \eqref{UVlimit} we infer 
the following deviation of the  solutions  at $k=0$ from their classical 
values for large values of $\vert\phi\vert$:
\begin{equation}
W_0'-W_{\Lambda}'\sim \frac{1}{2\phi},\quad Z_0'-Z_{\Lambda}'\sim \frac{1}{12\phi^4},\quad X_0-X_{\Lambda}\sim \frac{1}{18\phi^6},\quad  Y_0-Y_{\Lambda}\sim -\frac{1}{9\phi^7}\,.
 \label{largephi} 
 \end{equation}
As expected, the higher-order operators show a faster decay for large
field values, see Fig. \ref{fig:Flows1} and \ref{fig:Flows2}.

\subsection{Supersymmetry breaking}
\label{sec:SUSYBreaking}
If we choose the classical superpotential to be a polynomial of the form
$W_{\Lambda}'(\phi)\sim\mathcal{O}\left(\phi^{n}\right)$ with leading
power $n$ even, we expect spontaneous supersymmetry breaking to occur during
the flow towards the IR \cite{Wozar:2011gu, Synatschke:2010ub,  Witten:1981nf}.
It is known that spontaneous supersymmetry breaking is an IR effect, where the
ground state is lifted to $E_0>0$ \cite{Dine:2010cv}.
\subsubsection{Problems with the expansion in powers of $F$}
\label{subsubsec:Basics}
 In order to study SUSY breaking within the FRG framework we focus on the
$\mathbb{Z}_2$ symmetric even function
\begin{equation}
W_{\Lambda}'(\phi)=e+g \phi^2, \quad \, e<0,\; g>0.
\label{eq:Wclass}
\end{equation}Then the RG flow preserves the symmetry and $W_{k}'(\phi)$ will remain
$\mathbb{Z}_2$ symmetric for all scales. \\
For unbroken supersymmetry we employed an expansion in the auxiliary field
around $F=0$ to derive the flow equations in terms of the scalar fields $\phi$.
However, this expansion point is inappropriate when supersymmetry is broken 
in which case the vacuum expectation value of the average
field $F$ does not vanish. The problem with expanding around $F=0$ 
can be seen already in the LPA where $W''(\phi)$
represents a ''mass term'' in the denominator of the flow equation. Hence, the
regulator $r_1$ does not regulate since $\mathcal W''(\phi)=W''(\phi)+k$ 
will vanish for some value(s) of $\phi$.
This means that the RG equation detects the massless
fermionic excitation - the Goldstino mode - associated with the spontaneous breaking of
supersymmetry. This mode mediates between the two degenerate ground
states at $E_0>0$, one  in the bosonic and one in the fermionic sector
\cite{Wozar:2011gu}.
Hence, at the minimum of $V(\phi)$ the denominator in the flow equation simply
contains the squared Goldstino mass $m_G^2=W''(0)^2=0$.
Thus, the flow of the superpotential diverges in the IR limit at the origin.
This apparently leads to infinitely large excited energies, since
$E_1=W'(0)W^{(3)}(0)>0$ for broken supersymmetry. We find that this divergence
occurs independently of the choice of the regulator $r_2$ and of the 
order of truncation.\footnote{Of course, this IR problem represents a low
dimensional issue as the divergences diminish with increasing dimension $d$,
see \cite{Synatschke:2009nm, Synatschke:2010ub, Synatschke:2009da}.}\\
Thus, we are lead to Taylor-expand in powers of $F-F_0$ with
non-vanishing $F_0$. We shall do this in NLO in the derivative expansion. 
First we consider the equation of motion for the auxiliary field in  NLO, given by
\begin{align}
 F=-i W_k'(\phi)/Z_k'(\phi)^2.
\end{align}
If supersymmetry is spontaneously broken, $W_k'(\phi)>0$ for all $\phi$.
Again we observe that $F$ assumes a finite vacuum expectation value implying a 
breakdown of the flow equation when $W'$ ceases to have a zero. Now we expand 
around a non-zero auxiliary field - determined by its equation of motion - 
and rewrite the lhs of the flow equation as
\begin{align}
 \mathrm{lhs}&=iF\partial_kW'+\frac{1}{2}(\partial_k Z'^2) F^2+\mathcal{O}(F^3) \notag\\
&=\frac{W'\left(Z'\partial_k{W'}-W'\partial_k Z'\right)}{
Z'^3}+i\left(\partial_k{W'}-\frac{2 W'\partial_k Z'}{Z'}\right)(F-F_0)
+Z'\partial_k(Z')(F-F_0)^2+\dots\notag\\ 
&\;\text{with}\ \ F_0(\phi)=-i
W'(\phi)/Z'(\phi)^2.
\end{align}
Obviously the term $\mathcal{O}(F^3)$ will contribute to all orders around this
new expansion point. Unfortunately, there is no unique projection onto the 
flows of $W'$ and $Z'$. We may project onto the constant, the linear or the
quadratic term in $(F-F_0)$. Hence, the system is overdetermined. Note that
higher order terms contain no information about the flows of $W'$ and $Z'$.
Solving all three equations using an expansion of the rhs of the flow equation
yields no consistent solution, since  higher derivative
operators  contribute to these lower orders as well. To obtain  a
maximally self-consistent truncation it is therefore necessary to minimize
these contributions.  Assuming a nice convergence behaviour of the derivative
expansion, it is sensible to project onto the lowest orders in $(F-F_0)$.

\subsubsection{Numerical results}
\label{subsubsec:NumResultsSB}  

\begin{figure*}[ht]
\centering{ 
\includegraphics[width=0.49\textwidth]{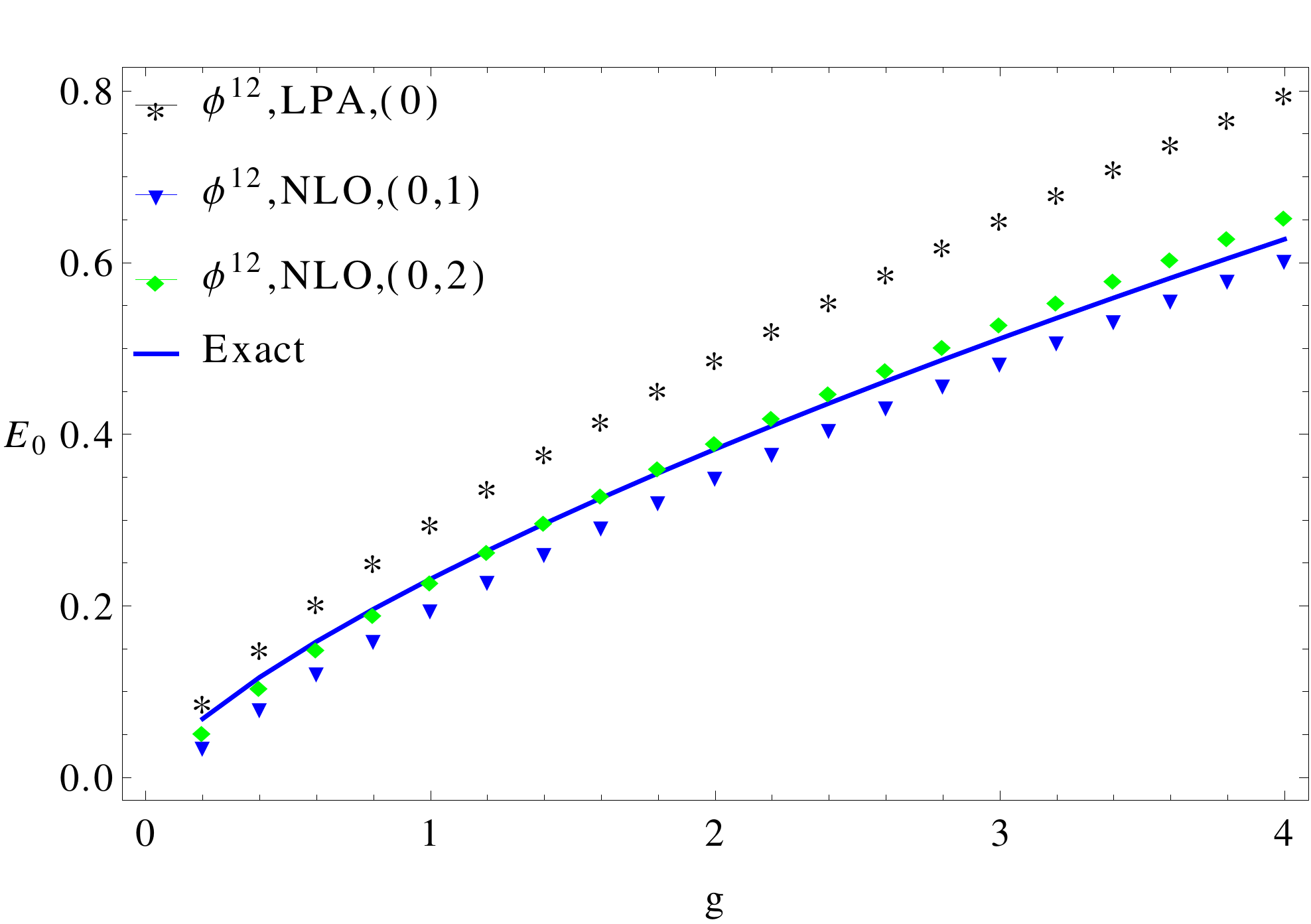}\hfill 
\includegraphics[width=0.5\textwidth]{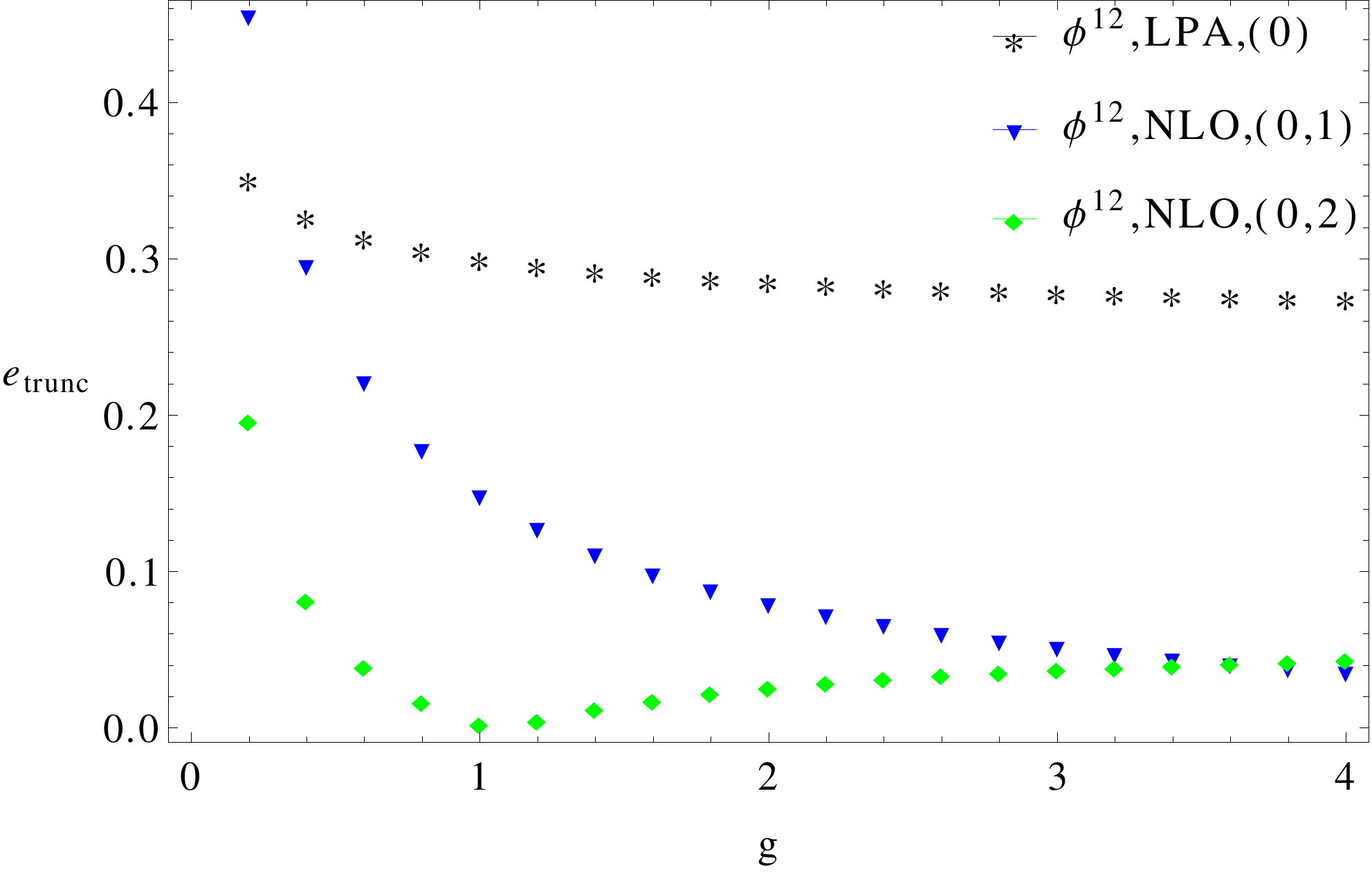}}
\caption{\label{brokenergs} Ground state energy $E_0$ and its relative error $e_{\mathrm{trunc}}$ 
 for initial potentials of the form $W_{\Lambda}=-0.1+\frac{g}{3}\phi^3$ as a
 function of $g$ obtained via a polynomial expansion of $W_k'(\phi)$ up to
 $\phi^{12}$. The brackets encode the projection scheme, i.e. $(i,j)$
 corresponds to a projection onto  $(F-F_0)^i$ and $(F-F_0)^j$.}
 \end{figure*}

In order to solve the flow equations for $W'$ and $Z'$ we
now limit our discussion to the approximation of a uniform wave-function
renormalization by setting
$Z_k'(\phi)=Z_k'$. This corresponds to neglecting the field- and momentum
dependence of $Z'$.  Then, the rhs of the RG equation \eqref{eq:wetterich}
simplifies to
\begin{align}
\mathrm{rhs}=& \frac{1}{2}\int\limits_{-k}^k
\frac{\mathrm{dq}}{2\pi}\Big(\partial_k(Z'^2 r_2)\big(\mathcal{W}''^2-B^2q^2\big)-2
(\partial_k{r_1}) B\mathcal{W}''\Big)
\left[\frac{ W'W'''}{\mathcal{N}(\mathcal{N}Z'^2+BW'W''')}
(F-F_0)^0\right.\notag\\
&\left.+\frac{iW'''Z'^4}{(\mathcal{N}Z'^2+BW'W''')^2}(F-F_0)^1+\frac{B(W''')^2Z'^6}{(\mathcal{N}Z'^2+BW'W''')^3}(F-F_0)^2+\dots\right]\\
&\text{with}\quad \mathcal N=B^2q^2+\mathcal{W}''^2,\quad\;
B=Z'^2(1+r_2),\quad\; \mathcal{W}''=W''+r_1.\notag
\end{align}
Note that we identified the wave-function renormalization
$Z'(\bar{\phi})$ which accompanies the regulator function $r_2$
with the field-independent $Z'$. To analyze the flow of
the effective average potential we proceed in two steps.  First we start
with a classical potential  of the form \eqref{potBreaking} in the UV at
$k=\Lambda$. Down to some scale $k_0>0$, $W'$ will have a zero. In this
regime $k\in(k_0,\Lambda)$ we employ the flow equations obtained by
an expansion around $F=0$.  
Starting with $k_0$  down to the IR limit $k=0$ the scale-dependence of $W',Z'$
is determined by the flow equations derived via an expansion around $F_0\neq 0$.
As regulator functions we choose $r_1=0$ and
$r_2=\left(k^2/p^2-1\right)\theta(k^2-p^2)$. 
To calculate the ground state energies $E_0$ we Taylor-expand
$W'$ about $\phi=0$ up to some order and solve the system
of  coupled ordinary differential equations numerically. This is a sensible 
approach when $W'$ becomes flat in the IR, because
due to supersymmetry the physics happens at vanishing field, in contrast to e.g. usual
$O(N)$-models \cite{Heilmann:2012yf,Litim:2011bf}, where the situation is
exactly opposite: in the unbroken regime, the derivative of the potential is positive, whereas in the broken phase, one has a zero.
As in the case of unbroken supersymmetry, we compare our results for $E_0$  with the ones obtained
by numerically diagonalizing the Hamiltonian of the system. \\ 
Fig. \ref{brokenergs} displays the ground state energies as well as the 
relative error
$e_{\text{trunc}}$ in
LPA and NLO as obtained via two different projection methods.   Here,  $(i,j)$
corresponds to a projection onto $(F-F_0)^i$ and $(F-F_0)^j$.
Apparently, the results are significantly improved by including a constant
wave-function renormalization. In particular, this applies to  large couplings
$g$, where the relative error is approximately $4\%$. Contrary to unbroken supersymmetry, the relative error increases with decreasing $g$. This originates 
from the fact that for decreasing $g$ the minima
of the potential drift apart and tunneling effects become effective, see
\cite{Zappala:2001nv}.\\ In NLO, a $(0,2)$-projection shows a smaller
relative error than the $(0,1)$-projection up to some $g_{\mathrm{max}}\approx 3.6$, since the flow of $Z'$ slows down when including the
higher order term $(F-F_0)^2$ resulting in a higher ground state energy
$E_0=V(0)=W'(0)/Z'(0)^2$. However, for large $g>g_{\mathrm{max}}$ the $(0,1)$-projection leads to superior results. This may be due to a larger truncation
error in $(F-F_0)^2$ compared to $(F-F_0)^1$ with increasing coupling $g$,
originating from the missing higher order terms $X,Y$ which are of importance there.
 
\section{$\mathcal{N}=1$ Wess-Zumino model in $3$ dimensions}
\label{sec:Wess-Zumino}

As a second testing ground for the convergence properties of the
supercovariant derivative expansion we choose the three-dimensional
$\mathcal{N}=1$ Wess-Zumino model. This model has been examined in next-to-leading 
order in the derivative expansion with a momentum- and field-independent wave-function 
renormalization $Z_k$ in \cite{Synatschke:2010ub}. It was shown that at zero 
temperature, this model possesses an analogue of the Wilson-Fisher fixed point, 
separating the supersymmetric (spontaneously broken $\mathbb{Z}_2$) from  the
nonsupersymmetric ($\mathbb{Z}_2$-symmetric) phase.

\subsection{Preliminaries}
\label{subsec:fundamentals}

Here, we shortly recall the main properties of the three-dimensional model. For more 
details in the context of flow equations  we refer to \cite{Synatschke:2010ub}.
The real scalar field $\phi$, real auxiliary field $F$ and
real fermion field $\psi$ are components of a real superfield
\begin{equation}
\Phi(x,\theta)=\phi(x)+\bar{\theta}\psi(x)+\frac{1}{2}\bar{\theta}\theta F(x).
\end{equation}
The supersymmetry variations $\delta_\epsilon\Phi$ are generated by fermionic 
supercharges $\cQ,\bar{\cQ}$, where $\cQ=-i\partial_{\bar\theta}-\slashed{\partial}\theta$ and
$\bar{\cQ}=-i\partial_{\theta}-\bar{\theta}\slashed{\partial}$ with
anticommutation relations
$\{\cQ_k,\bar{\cQ}_l\}=2i\slashed{\partial}_{kl}$. The supercovariant derivatives - 
fulfilling the corresponding relation $\{D_k,\bar{D}_l\}=-2i\slashed{\partial}_{kl}$ - read
\begin{equation}
D=\partial_{\bar\theta}+i\slashed{\partial}\theta, \quad\mbox{and}\quad
\bar{D}=-\partial_{\theta}-i\bar{\theta}\slashed{\partial}.
\end{equation}
Since there are no Majorana fermions in $3$ Euclidean dimensions
we switch to Minkowski-space with metric tensor $\eta_{\mu\nu}=\mbox{diag}(1,-1,-1)$ 
and $\gamma$-matrices $\gamma^{\mu}=(\sigma_2, i\sigma_3, i \sigma_1)$, 
where $\mu=0,1,2$. After setting up the flow equation we return to
Euclidean space, see \cite{Synatschke:2010ub}.
With the above definitions we are able to construct the off-shell
supersymmetric action in $\mathbb{R}^{1,2}$ superspace:
\begin{equation}
S[\Phi]=\int dz
\left[-\frac{1}{2}\Phi K \Phi+2W(\Phi)\right], \quad
K=\frac{1}{2}(\bar{D}D-D\bar{D}),
\end{equation}
where $z=(x, \theta, \bar\theta)$ denotes the coordinates in superspace.
After integration over the Majorana Grassmann variables $\theta, \bar\theta$ and 
elimination of the auxiliary field $F$ via its equation of motion $F=-W'(\phi)$, 
we arrive at the following on-shell action in components,
\begin{equation}
S_{on}[\phi, F, \psi, \bar\psi]=\int d^3x\left[\frac{1}{2}\partial_\mu\phi
\partial^\mu \phi-\frac{i}{2}\bar\psi \slashed{\partial}\psi
-\frac{1}{2}W'^2(\phi)-\frac{1}{2}W''(\phi)\bar\psi \psi\right].
\label{eq:SonWess} 
\end{equation} 
The last term in (\,\ref{eq:SonWess}) describes a Yukawa interaction
between the scalars and fermions and
$V(\phi)=W'{}^2(\phi)/2$ the potential self-energy of the scalars.

\subsection{Derivation of flow equation}
\label{subsec:deriavtionWess-Zumino} 

Now we use the flow equation in Minkowski spacetime
\begin{equation}
\partial_k\Gamma_k=
 \frac{i}{2} \STr\left\{\partial_k  R_k\left[\Gamma_k^{(2)}+
 R_k\right]^{-1}\right\}
\label{eq:wetterichMinkowski}
\end{equation}
and perform a
Wick rotation of the zeroth momentum component afterwards to obtain the corresponding
flow in Euclidean space.\\ Analogously to 
eq.\,(\ref{flow0}), the general ansatz for the scale-dependent effective average action reads
\begin{equation}
\Gamma_k[\Phi]= \int \text{d}z \left[2\, W_k(\Phi)-\frac12 Z_k(\Phi)
    K Z_k(\Phi)-\frac{1}{8}Y_{1,k}(\Phi)K^2\Phi-\frac{1}{8}Y_{2,k}(\Phi)(K\Phi)(K\Phi)\right]
    \label{GammakWess}
\end{equation}
with the scale- and field-dependent functions $W_k, Z_k,Y_{1,k}, Y_{2,k}$.
We chose the prefactors of each term such that the resulting flow
equations in Euclidean space exactly match the corresponding flow equations derived in
supersymmetric quantum mechanics with $\int
\frac{dq}{2\pi}\rightarrow\int\frac{d^3q}{(2\pi)^3}$. By integrating over
the anticommuting Grassmann variables in (\ref{GammakWess}) we get
\begin{align}
&\Gamma_k[\Phi]=\!\!\int\!\!
d^3x\left[\frac{1}{2}(\partial_\mu Z)(\partial^\mu
Z)-\frac{i}{2}(Z'\bar\psi)\slashed\partial(Z'\psi)-\frac{1}{4}Y_1'\bar\psi\partial^2\psi-\left(\frac{1}{2}W''+\frac{1}{8}Y_1''(\partial^2\phi)\right)\bar\psi\psi\right.\notag\\
&+\frac{1}{4}Y_2(\partial^\mu\bar\psi)(\partial_\mu\psi)
+\left(W'-\frac{1}{2}Z'Z''\bar\psi\psi+\frac{1}{2}(Y_1'+Y_2)(\partial^2\phi)+\frac{1}{4}Y_1''(\partial_\mu\phi)(\partial^\mu\phi)+\frac{i}{2}Y_2'\bar\psi\slashed\partial\psi\right)F\notag\\
&\left.+\left(\frac{1}{2}Z'^2+\frac{1}{8}Y_2''\bar\psi\psi\right)F^2-\frac{1}{4}Y_2'F^3\right],
\label{GammaWesscomponents}
\end{align}
where again we ordered the terms in powers of the auxiliary field $F$.
In analogy to (\ref{reg3}), we assume the supersymmetric cutoff action to be of
the form
\begin{equation}
\Delta S_k[\Phi] = 
\frac12  \int \text{d}z\, \Phi\left[
2r_1(-\partial_\mu\partial^\mu,k)-Z_k'{}^2(\bar{\Phi})\,r_2(-\partial_\mu\partial^\mu,k)K\right]\Phi,
 \label{regWess}
\end{equation}
with $Z_k'$ evaluated at the background field $\bar\Phi=\bar\phi$.
As in quantum mechanics, we extract the scale
dependence of $W',Z'$ and $Y_2'$ by projecting the rhs of
(\ref{eq:wetterichMinkowski}) onto $F,F^2$ and $F^3$ for constant
bosonic fields and a vanishing fermion field.
This way we obtain (cf. eq.\,(\ref{fluss1}))
\begin{align}
&\left.\partial_k
\Gamma_k\right|_{\partial_\mu\phi=\partial_\mu F=\psi=\bar{\psi}=0}= \int
\text{d}^3 x \left( \partial_k W'F+\frac{1}{2}\partial_kZ'^2
F^2-\frac{1}{4}\partial_kY_2' F^3\right)\notag\\ 
&=\frac{i}{2}\int
\frac{\text{d}^3q}{(2\pi)^3}\,\frac{\text{d}^3q'}{(2\pi)^3}\,\text{d}\theta
\,\text{d}\bar{\theta}\,\text{d}\theta'\,\text{d}\bar{\theta}'(\partial_kR_k)(q',q,\theta',\theta)G_k(q,q',\theta,\theta')\notag\\
&=\frac{i}{2}\int
\text{d}^3x\,\frac{\text{d}^3q}{(2\pi)^3}\left[(2\partial_kr_1)(b+c+d)+\partial_k(r_2Z_k'^2(\bar{\phi}))(-2f
q^2+a q^2+4e)\right],
\label{flussWess1}
\end{align} 
where $(a,b,c,d,e,f)$ again represent the coefficients of the Greens function
$G_k$, see appendix \ref{sec:appendix4}.
Finally, we obtain the flow equations in Euclidean space 
with metric $-\delta_{\mu\nu}$ via a Wick rotation of
the zeroth momentum component, i.e. $q^0 \rightarrow i q^0$. 
As mentioned above, these equations are by construction identically 
to the ones derived in SUSY quantum mechanics up to an integration 
over a three dimensional momentum space.\\
The missing flow of $Y_1'+Y_2$ is derived in an exactly similar manner to
$d=1$ as presented in appendix \ref{sec:appendix2} by considering momentum-dependent fields $\phi,F$ and projecting onto the contribution in $\cQ^2
\delta\phi(\cQ)\delta F(-\cQ)$  with an additional Wick rotation of $q^0$
afterwards. Finally, we define and substitute $Y:=Y_2'$ and
$X:=Y_1'+Y_2$ in order to simplify the equations.

\subsection{Superscaling relation \& Wilson-Fisher fixed point}
\label{subsec:WFfixedpoint}

\begin{figure*}[ht]
\centering{
\includegraphics[width=0.98\textwidth]{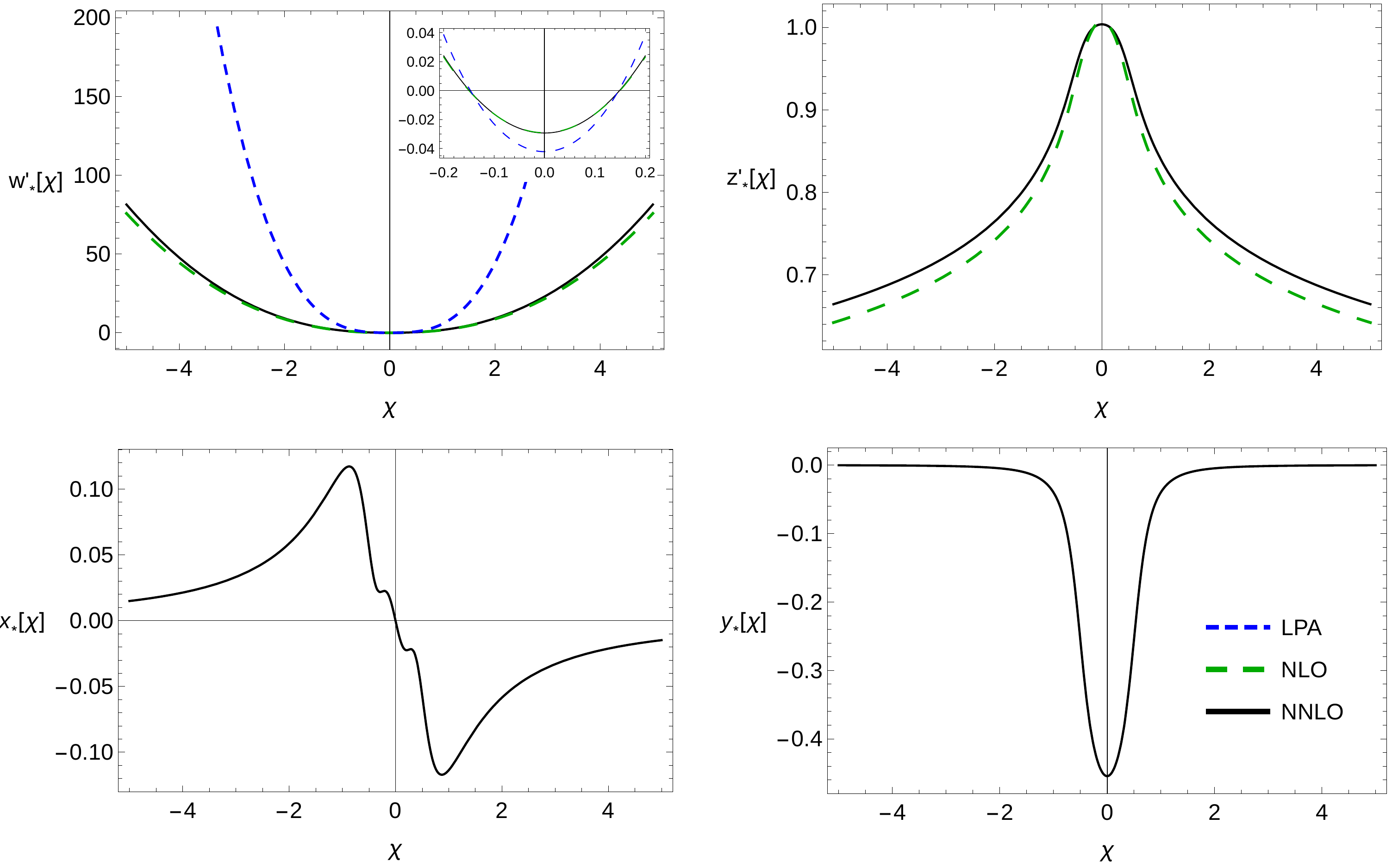}
}
\caption{Fixed-point solutions of the three-dimensional Wess-Zumino model in LPA, NLO and NNLO. The change in asymptotics when going from LPA to NLO induced by
the anomalous dimension is clearly visible. Also, the potential as well as the
wave function renormalization change only  mildly when going from NLO to NNLO,
indicating once more a good convergence of the derivative expansion.}
\label{fig:FPfunctionals}
\end{figure*}

In this section we study the fixed-point structure  of
the three-dimensional Wess-Zumino model in NNLO. To begin with, we 
have to choose suitable regulator functions.
At the cutoff scale we assume $W'_\Lambda$ to be an even polynomial.
Then its leading power remains even during the flow and we can investigate
the dynamical breaking of supersymmetry during the flow to the IR. 
The flow equations automatically force $W'(\phi)$ to remain even for 
all scales $k<\Lambda$. Furthermore, the
equations then imply that $Z'(\phi)$ and $Y(\phi)$ are even and $X(\phi)$ is
odd.
If $W'(\phi)$ tends asymptotically to an even power, 
$W''(\phi)$ always has a node which is merely shifted but not screened by
the mass-like regulator $r_1$. Thus, we may set $r_1=0$. We select an optimized
regulator function $r_2$ of the form\footnote{The choice (\ref{eq:regWess}) with $n=1$
as used in \cite{Synatschke:2010ub} would lead to IR divergent flows of the forth-order operators $X$ and $Y$.}
\begin{equation}
r_2(q^2,k)=\left(\frac{k^n}{|q|^n}-1\right)\Theta\left(\frac{k^2}{q^2}-1\right)\quad
\mbox{with}\quad n=2.\label{eq:regWess}
\end{equation}
We identify the value of the background field appearing in the
cutoff action with the minimum of the potential, i.e. $\bar{\phi}=\phi_0$. We
emphasize that we do not stick to the  background field approximation
\cite{Reuter:1993kw, Synatschke:2008pv} going along with the identification
$\bar{\phi}=\phi$ for reasons given later in this section.\\ Next, we switch to dimensionless quantities in order to analyze the characteristics of the phase transition in $d=3$. 
With the canonical dimensions
\begin{equation}
[\phi]=1/2,\quad [W]=2,\quad [Z']=0,\quad [X]=-1,\quad [Y]=-3/2\, ,
\end{equation}
we are lead to the transformations
\begin{align}
\chi&=Z'(\phi_0)k^{-1/2}\phi,\quad\quad\quad
w(\chi)=W(\phi)k^{-2},\quad\quad z'(\chi)=Z'(\phi)/Z'(\phi_0)\, ,\notag\\
 x(\chi)&=X(\phi)k/Z'^{2}(\phi_0),\quad\quad\,\,\,
y(\chi)=Y(\phi)k^{3/2}/Z'^{3}(\phi_0)\, ,
\label{Wessdeimanionless}
\end{align} 
with $\phi_0$ denoting the minimum of the potential.
Employing the definition of the anomalous dimension 
\begin{equation}
\eta(k)=-\frac{d}{dt}\ln \left(Z'^{\,2}(\phi_0)\right),\quad\, t=\ln(k/\Lambda)\, ,
\label{eq:anomalous}
\end{equation}
as well as the dimensionless momentum variable $u=q^2/k^2$, the flow of the
dimensionless superpotential reads
\begin{align}
&\partial_t w'+\frac{1}{2} (3-\eta) w'-\frac{1}{2} (\eta +1) \chi 
w''=\notag\\
&\frac{1}{2\pi^2}\int_{0}^{1}
\text{d}u\frac{\sqrt{u} (\eta  (u-1)+2) \left(\alpha '(u \alpha ^2  -4
\beta^2) +16 \alpha  u z' z'' \beta\right)}{2 \left(u \alpha^2 +4
\beta^2\right)^2},
\label{flowwWess}
\end{align}
where we abbreviated $\alpha:=u x+2 w''$ and $\beta=1+u(z'^2-1)$.  The lhs of
(\ref{flowwWess}) includes the dimensional and anomalous scaling, whereas the
rhs  encodes the interactions amongst the operators according to our ansatz of $\Gamma_k$.
Note that of the fourth-order contributions only $x$ and not $y$ directly couples to the flow of
the superpotential. The expressions of the remaining flows
are rather long and therefore not written down explicitly.\\
Now we analyze the physics at the  Wilson-Fisher fixed point, corresponding to
\begin{equation}
\partial_t \mathcal{O}_*=0,\quad\, \mbox{with} \quad\, \mathcal{O}=(w',z',x,y).
\label{fixedpoint}
\end{equation}
For large fields  $|\chi|\gg 1$ the rhs of eq.\,(\ref{flowwWess}), i.e. the
nontrivial flow, vanishes as  we generally expect $|w_*''|$ to be
 large  for a $\mathbb{Z}_2$-symmetric system. This holds for
the remaining flows  as well. Thus, the fixed-point solution
for large $\chi$ is fixed by the anomalous and canonical scaling, leading to
the asymptotic behaviour
\begin{align}
w_*'(\chi)\sim \chi^{\left(\frac{3-\eta}{\eta+1}\right)},\quad z_*'(\chi)\sim
\chi^{-\left(\frac{\eta}{\eta+1}\right)},\quad x_*(\chi)\sim \chi^{-2},\quad
y_*(\chi)\sim \chi^{-3}.
\label{scalinglargefields}
\end{align}
Thus, for positive $\eta$, the higher order functions vanish for large fields.
In \cite{Synatschke:2010ub}, one non-Gaussian IR-stable
fixed point as the supersymmetric analogue of the Wilson-Fisher fixed point 
has been spotted. It possesses
one IR unstable direction  defined by $w_*'(0)$ with critical exponent
$\theta_0=1/\nu_\text{w}$ which can be related to the anomalous dimension $\eta$
via the \textit{superscaling relation}  $1/\nu_{\text{w}}=(3-\eta)/2$ 
\cite{Synatschke:2010ub}.\\ 
To begin with, we show that this interesting superscaling relation holds true 
to all orders in the supercovariant derivative expansion of $\Gamma_k$ and 
derive its form for arbitrary $d \geq 2$. \\
To show this, let us note that the only fixed point equation which depends
explicitly on $w'$ is the one for $w'$ itself\footnote{Note that in the
subsequent derivation we assume $\eta$ to be independent of a scale-dependent
$\phi_0$. We expect corrections due to the running of the VEV to be small, which
is confirmed by our numerical studies.}. Thus, we consider small
fluctuations around the fixed-point solution in $w'$-direction, i.e. $w'(t,\chi)= w'_*(\chi) + \delta w'(t,\chi)$ and possible higher order operators evaluated at the fixed point. By linearizing the flow \eqref{flowwWess} - generalized to $d$ dimensions\footnote{The rhs of \eqref{flowwWess} holds for arbitrary $d$ up to a different dimensional prefactor of $1/(2^{d-1}\pi^{d/2}\Gamma(d/2))$.} - in
 $\delta w'$ we arrive at the fluctuation equation
\begin{equation}
 \partial_t \delta w' = \left(\frac{\eta-d}{2} + \mathcal{F}(\chi)
 \partial_\chi + \mathcal{G}(\chi) \partial_\chi^2 \right) \delta w' \, .
\end{equation}
Here, $\mathcal{F}$ and $\mathcal{G}$ are functionals obtained from the linearization.
The critical exponents then correspond to the  negative eigenvalues of the
operator on the rhs. Apparently, a constant variation is an
eigenfunction to this operator with eigenvalue $(\eta-d)/2$. Since the flow
of all higher operators of $\Gamma_k$ remain independent of $w'$ this is true to
all orders. Hence, we have verified the superscaling relation
\begin{equation}
\frac{1}{\nu_\mathrm{w}}=\frac{1}{2}(d-\eta), \quad \quad d\geq2.
\end{equation}
Next we present the numerical results to the fixed point equations. We solved
the fixed point equation globally via  a combination of Chebyshev and rational
Chebyshev polynomials. The fixed-point solutions of the four operators
considered are illustrated  in Fig. \ref{fig:FPfunctionals}. The IR relevant
coupling $w'(0)$, the location of the minimum of the potential as well as the
anomalous dimension and  leading critical exponents are displayed in Table
\ref{tab:FPvals}. We observe a nice convergence behaviour with increasing
order in the derivative expansion. This confirms that the quantitative relevance
of the operators in $\Gamma_k$ correlates with their scaling dimension $D$
(canonical plus anomalous  scaling). They are ordered as
\begin{equation}
D_{w'}>D_{z'}>D_x>D_y\, ,
\end{equation}
and are determined in terms of $\eta$ as follows:
\begin{equation}
D_{w'}=\frac{3}{2}-\frac{\eta}{2},\;\; D_{z'}=-\frac{\eta}{2},\;\;
D_x=-(1+\eta),\;\; D_y=-\frac{3}{2}(1+\eta).
\label{scalingdimension}
\end{equation}

\TABLE{
\begin{tabular}{l||lllllll} 
\hline \hline
approximation&
$w_*'(0)$&$\chi_0$&$\eta$&$\theta_0$&$\theta_1$&$\theta_2$&$\theta_3$\\
\hline
LPA&$-0.0420$&$0.147$&& $3/2$ & $-0.702$ & $-3.800$&$-7.747$\\
NLO&$-0.0292$&$0.150$&$0.186$&$1.407$&$-0.771$&$-1.642$&$-3.268$\\
NNLO&$-0.0294$&$0.149$&$0.180$&$1.410$&$-0.715$&$-1.490$&$-2.423$\\
\hline\hline 
\end{tabular}
\label{tab:FPvals}
\caption{Value of the superpotential $w_*'(\chi=0)$ at the origin, location of
minimum $\chi_0$ of the fixed-point potential, anomalous dimension and leading
and subleading critical exponents of the Wilson-Fisher fixed point for different
orders in the supercovariant derivative expansion.} 
}
Finally, we wish to comment on our decision not to use the background field
approximation (BFA) for the spectrally adjusted regulator $r_2$.
According to \eqref{scalinglargefields}, the  fixed point solution $z_*'$
vanishes for large fields $|\chi|\rightarrow \infty$. Implementing the BFA goes along
with the replacement $z'(\chi_0)=1\rightarrow z'(\chi)$ in the
dimensionless regulator. Thus, the regulator  is suppressed artificially for
large fields. This in turn can lead to instabilities.
Indeed, during our numerical investigation of the fixed point equations
we could not find a global solution in NNLO when employing the background 
field approximation, even though a solution via a Taylor expansion
seems to exist. The difference between the physical quantities (critical
exponents etc.) obtained by a Taylor expansion with BFA and the spectral method 
without BFA are almost identical. Thus one might argue that
the error made in this approximation is irrelevant. However, one should
bear in mind that a fixed point potential better be globally defined, and that there
might be systems that are unstable against such types of approximations. Note
also that when integrating the dimensionful flow equations, the difference
should be even smaller as $Z'(\phi)$ does not fall off asymptotically.

\newpage
\section{Summary}

In this paper we studied the convergence of the derivative expansion in two
supersymmetric theories via the functional renormalization group. Our approach
includes a manifestly supersymmetric regulator and thus is suited to study 
systems with broken and unbroken supersymmetry.

In the first part, we investigated  supersymmetric quantum mechanical systems. Starting with
symmetry-preserving flows, we obtained very good results for the energy gap
within a truncation containing forth-order derivative terms. The relative error
to the exact results is below one percent for a wide range of couplings
$0<g<\sqrt3$ including the non-perturbative regime. For larger couplings, we
observe a breakdown of the derivative expansion, which is expected since in this regime non-local instanton contributions play a crucial role. In the SUSY-breaking case, we studied the flow of the superpotential together
with a constant wave-function renormalization. This required a suitable choice
of projection to obtain the flow equations, because in the broken regime, the
auxiliary field acquaints a finite expectation value which has to be accounted for.
The results are again in agreement to exact results within a few percent.

The second part of this work deals with the Wess-Zumino model in 3 dimensions.
In particular, we were interested in the analogue of the Wilson-Fisher fixed
point of standard $O(N)$-theories. We were able to calculate all functionals 
at the fixed point up to forth order in supercovariant derivatives.
Again, a very good convergence has been observed, and this convergence
is further substantiated by two facts: 
On the one hand, the change of the fixed functionals is tiny when including
forth-order terms in the effective average action. On the other hand, the
critical exponents converge. Additionally, we could
prove the superscaling relation to all orders in the derivative expansion.

In both studies, globally defined spectral methods were used to obtain
the results. This ensures the numerical validity of our results, which are naturally
free from any boundary effects usually present when applying a domain truncation.

\acknowledgments{The authors want to thank J. Borchardt, J.M. Pawlowski,
 R. Sondenheimer, F. Synatschke-Czerwonka
and L. Zambelli
for useful discussions. The research was supported by the Deutsche
Forschungsgemeinschaft (DFG) graduate school GRK 1523/2.
B. Knorr and A. Wipf thank the DFG for supporting this work under
grant no. Wi777/11-1.}
\newpage
\appendix   
\section{Greens function in $d=1$}
\label{sec:appendix1}
An analysis  of the flow equation \eqref{eq:wetterich} requires the determination of the connected two-point function $G_k$ as the inverse of $(\Gamma^{(2)}_k+R_k)$. 
In superspace, the relation
$\mathbbm{1}=G_k(\Gamma^{(2)}+R_k)$ is given by \eqref{defGreen} with $G_k$ 
according to \eqref{ansatz}. The latter simply represents an expansion of $G_k$ in the Grassmann variables $(\theta, \theta^{\prime})$ with arbitrary ``bosonic'' coefficients $(a,b,c,d,e,f)$.
By solving \eqref{defGreen}, we finally obtain 
\begin{align}
a&=\frac{\left(B+\frac{3}{2}i F Y_2'\right)}{ \frac{1}{16} \left(4 A+F \left(3 F Y_2''-8 i Z' Z''\right)\right){}^2+\left(B+\frac{3}{2}i F Y_2'\right) C}\notag\\
b&=c=\frac{ -\left(i A+\frac{3}{4} i F^2 Y_2''+2 F Z' Z''\right)}{ \left(\frac{1}{16} \left(4 A+F \left(3 F Y_2''-8 i Z'
   Z''\right)\right){}^2+\left(B+\frac{3}{2}i F Y_2'\right) C\right)}\notag\\
d&=\frac{4 i}{4 A-4 i B q+F \left(F Y_2''+4 q Y_2'-4 i Z' Z''\right)}\notag\\
e&=\frac{4 i}{4 A+4 i B q+F \left(F Y_2''-4 q Y_2'-4 i Z' Z''\right)}\label{coefficients}\\
f&=\frac{1}{\left(B+\frac{3}{2}i F Y_2'\right)}+\frac{\left(i A+\frac{3}{4} i F^2 Y_2''+2 F Z' Z''\right){}^2}{\left(B+\frac{3}{2}i F Y_2'\right) \left(\frac{1}{16} \left(4 A+3 F^2 Y_2''-8 i F Z' Z''\right){}^2+\left(B+\frac{3}{2}i F Y_2'\right) C\right)},
\notag
\end{align}
where we introduced the abbreviations
\begin{align}
 A&=W''+r_1+\frac{1}{2}(Y_1'+Y_2)q^2\notag\\
   B&=Z'^2+r_2 Z_k'^2(\bar{\phi})\notag\\
   C&=B q^2+\frac{i}{4}  F^3 Y_2{}'''+F^2 \left(Z''^2+Z''' 
   Z'\right)+\frac{i}{2} F \left(q^2 \left(2 Y_2'+Y_1''\right)+2 W'''\right).
\label{abbrev1}
\end{align}

\section{Flow equation of $Y_1'+Y_2$} 
\label{sec:appendix2}

This section explains the derivation of the flow of $Y_1'+Y_2$. To derive the corresponding flow equation, we project the rhs of \eqref{eq:wetterich} onto the time dependent term   $\dot{F}\dot{\phi}$.
 This requires an expansion of the inverse propagator around  field configurations of $F$ and $\phi$ exhibiting a small momentum dependence. Again, we therefore only consider the bosonic  part of the superfield in the inverse propagator, i.e. we set $\psi$ and $\bar{\psi}$ equal to zero. The background field configurations in momentum space are then given by
\begin{align}
\phi(p)&= \phi \delta(p)+ \delta\phi(p)\left(\delta(p-Q)+\delta(p+Q)\right) \quad \;\mbox{and}\notag\\
F(p)&= F \delta(p)+ \delta F(p)\left(\delta(p-Q)+\delta(p+Q)\right)
\label{momentum1}
\end{align}
with $\delta\phi(p)\ll\phi, \delta F(p)\ll F$. Note that  $\phi(p)=\phi^*(-p)$ is real and $F(p)=-F^{*}(-p)$ purely imaginary.
Next, we perform an expansion of the inverse Green's function $\Gamma^{(2)}_k+R_k$ 
up to $\mathcal{O}\left(\delta\phi(Q)\delta F(-Q)\right)$ quadratic in the fluctuations. In the following, $z$ denotes  the superspace coordinates $(q,\theta, \bar{\theta})$. Thus, the inverse propagator may be written  in the form
\begin{equation}
\left(\Gamma^{(2)}_k+R_k\right)(z,z^{\prime})=\left[\Gamma_0(q,q^{\prime})+\Gamma_\phi(q,q^{\prime})+\Gamma_F(q,q^{\prime})+\Gamma_{\phi F}(q,q^{\prime})\right]
\delta(\theta,\theta^{\prime}),
\label{expansion}
\end{equation}
corresponding to an expansion in powers of $\delta\phi$ and $\delta F$ with
 \begin{align}
 \Gamma_0(q,q^{\prime})&=\hat{\Gamma}_0(q)\delta(q-q^{\prime})\quad \mbox{with}\quad     \notag\\
  &\phantom{=,}\hat{\Gamma}_0(q)=\left(i
  (W''+r_1)-BK(q)+\frac{i}{2}(Y_1'+Y_2)q^2\right),\notag\\ 
  &\phantom{=,}\hat{\Gamma}^{-1}_0(q)=\frac{-i(W''+r_1+\frac{1}{2}(Y_1'+Y_2)q^2)-B
  K(q)}{B^2q^2+(W''+r_1+\frac{1}{2}(Y_1'+Y_2)q^2)^2}\notag\\
 \Gamma_\phi(q,q^{\prime})&=\hat{\Gamma}_\phi(q,Q)\delta(q-q^{\prime}-Q)+\hat{\Gamma}_\phi(q,-Q)\delta(q-q^{\prime}+Q)\quad\mbox{with}\quad\notag\\
 &\phantom{=,}\hat{\Gamma}_\phi(q,Q)=\left(i W'''+Z'Z''Q^2\bar{\theta}\theta-Z'Z''(K(q)+K(q-Q))-\frac{i}{2}Y_2'Q^2\bar{\theta}\theta K(q-Q)\right.\notag\\
 &\phantom{=,}\left.-\frac{i}{2}Y_2'Q^2K(q)\bar{\theta}\theta+\frac{i}{2}Y_2'K(q)K(q-Q)+\frac{i}{4}Y_1''(q^2+Q^2+(q-Q)^2)\right)\delta\phi(Q),\notag\\
 \Gamma_F(q,q^{\prime})&=\hat{\Gamma}_F(q,Q)\delta(q-q^{\prime}-Q)+\hat{\Gamma}_F(q,-Q)\delta(q-q^{\prime}+Q)\quad\mbox{with}\quad\notag\\
 &\phantom{=,}\hat{\Gamma}_F(q,Q)=\left(i
 W'''\bar{\theta}\theta+Z'Z''-Z'Z''K(q)\bar{\theta}\theta
 -Z'Z''\bar{\theta}\theta K(q-Q) -\frac{i}{2}Y_2'K(q)\right.\notag\\
  &\phantom{=,}\left.-\frac{i}{2}Y_2'K(q-Q)+\frac{i}{2}Y_2'K(q)\bar{\theta}\theta K(q-Q)+\frac{i}{4}Y_1''(q^2+Q^2+(q-Q)^2)\bar{\theta}\theta\right)\delta F(Q),\notag\\ \Gamma_{\phi F}(q,q^{\prime})&=\hat{\Gamma}_{\phi F}(q)\delta(q-q^{\prime})\quad \mbox{with}\quad\notag\\
 &\phantom{=,}\left.\hat{\Gamma}_{\phi F}(q)\right|_{\mathcal{O}(Q^2)}=\frac{i}{2}(Y_2''+Y_1''' )Q^2 \bar{\theta}\theta(\delta\phi(Q)\delta F(-Q)+\delta F(Q)\delta \phi(-Q)).\label{drei}
 \end{align}
Thereby, we considered only terms quadratic in $Q$ in  $\Gamma_{\phi F}(q,q^{\prime})$ as other terms do not contribute to the flow. Note that the operator $K$ occurring in eq.\,(\ref{drei}) is a function of the momentum as well as  the Grassmann variables $\bar{\theta},\theta$.
Inserting the configuration \eqref{momentum1} into \eqref{flow1}  we obtain
\begin{align}
\frac{i}{2}(Y_1'+Y_2)=\frac{1}{\Omega}\left.\lim_{Q^2\rightarrow
0}\frac{\partial}{\partial Q^2} \frac{\delta^2 \Gamma_k}{\delta(\delta
\phi(Q))\delta(\delta F(-Q))}\right|_{\phi,\, F=\psi=\bar{\psi}=\delta F=\delta
\phi=0},
\label{projection}
\end{align}
where $\Omega$ denotes the total volume of ``space'' and should be taken to infinity at the end.
Now, the Green's function follows from \eqref{expansion} via an expansion in powers of $\delta\phi$ and $\delta F$. Keeping only contributions quadratic in the fluctuations $\sim (\delta F)( \delta \phi)$ we find
\begin{align}
\left(G_k\right)(z,z^{\prime})_{(\delta\phi \delta F)}&=\left\{ -\hat{\Gamma}^{-1}_0(q)\Gamma_{\phi F}(q,q^{\prime})\hat{\Gamma}^{-1}_0(q)\right.\notag\\
&+\int_{\tilde{q}}\hat{\Gamma}_{0}^{-1}(q)\Gamma_\phi(q,\tilde{q})\hat{\Gamma}_0^{-1}(\tilde{q})\Gamma_F(\tilde{q},q^{\prime})\hat{\Gamma}_0^{-1}(q)\notag\\
&\left.+\int_{\tilde{q}}\hat{\Gamma}_{0}^{-1}(q)\Gamma_F(q,\tilde{q})\hat{\Gamma}_0^{-1}(\tilde{q})\Gamma_\phi(\tilde{q},q^{\prime})\hat{\Gamma}_0^{-1}(q)\right\}\delta(\theta,\theta^{\prime})
\label{Greens}
\end{align}
Inserting the above propagator (\ref{Greens}) into the flow equation  and
keeping only terms in $\delta\phi(Q)\delta F(-Q)$ we have
\begin{align}
&\frac{i}{2}\partial_k(Y_1'+Y_2)\,\,Q^2\,\delta\phi(Q)\delta F(-Q)=\frac{1}{2}\int\frac{\text{d}q}{2\pi} \text{d}\theta  \text{d}\bar{\theta} \text{d}\theta'\text{d}\bar{\theta}'\partial_k\left[i r_1 -Z_k'^2(\bar{\phi}) r_2 K(q,\theta')\right]\delta(\theta', \theta)\times\notag\\
&\left[-\hat{\Gamma}^{-1}_0(q)\hat{\Gamma}_{\phi F}(q)\hat{\Gamma}^{-1}_0(q)+ \hat{\Gamma}^{-1}_0(q)\hat{\Gamma}_\phi(q,Q)\hat{\Gamma}_0^{-1}(q-Q) \hat{\Gamma}_F(q-Q,-Q)\hat{\Gamma}^{-1}_0(q)\right.\notag\\ 
&+ \left.\hat{\Gamma}^{-1}_0(q)\hat{\Gamma}_F(q,-Q)\hat{\Gamma}_0^{-1}(q+Q) \hat{\Gamma}_\phi(q+Q,Q)\hat{\Gamma}^{-1}_0(q)\right]\delta(\theta,\theta^{\prime}).
\label{ableit1} 
\end{align} 
By  integrating over the Grassmann variables and projecting the rhs onto the contribution $\sim Q^2$ we thus get the flow of $Y_1'+Y_2$.

\section{Greens function in $d=3$}
\label{sec:appendix4}
Similarly to the analysis in $d=1$, we determine the Greens function $G_k$ of
the general form
\begin{equation}
G_k(q,q',\theta,\theta')=\left(a+b\,\bar{\theta}\theta+c\,\bar{\theta}'\theta'+d\,\bar{\theta}\theta'+e\,
\bar{\theta}\theta\bar{\theta}'\theta'+f\bar{\theta}'\!\slashed{p}\theta\right)\delta(q,q')
\label{ansatzWess} 
\end{equation}
by solving
\begin{equation}
\int  \frac{\text{d}^3q'}{(2\pi)^3} \,\text{d}\theta'\,
\text{d}\bar{\theta}'\,G_k^{-1}(q,q',\theta,
\theta')\,G_k(q',q'',\theta',\theta'')=\delta(q,q'')\delta(\theta, \theta''),
\label{defGreenWess}
\end{equation}
for the coefficients $(a,b,c,d,e,f)$. This yields 
\begin{align}
a&=\frac{8 \left(2 B-3 F Y_2'\right)}{8 C \left(2 B-3 F
Y_2'\right)-\left(4 A-3 F^2 Y_2''+8 F Z' Z''\right)^2}\notag\\
b&=c= \frac{8 A-6 F^2 Y_2''+16 F Z' Z''}{\left(4 A-3 F^2 Y_2''+8 F Z'
Z''\right){}^2+8 C \left(3 F Y_2'-2 B\right)} \notag\\
d&=-\frac{4 \left(A-\frac{1}{4} F^2 Y_2''+F Z' Z''\right)}{4 \left(A-\frac{1}{4}
F^2 Y_2''+F Z' Z''\right){}^2-4 q^2 \left(B-F Y_2'\right){}^2}\notag\\
e&=\frac{4 C}{8 C \left(2 B-3 F Y_2'\right)-\left(4 A-3 F^2 Y_2''+8 F Z' Z''\right){}^2}\notag\\
f&=-\frac{16 \left(B-F Y_2'\right)}{\left(4 A-4 B q+F \left(-F Y_2''+4 q Y_2'+4
Z' Z''\right)\right) \left(4 A+4 B q-F \left(F Y_2''+4 q Y_2'-4 Z'
Z''\right)\right)},
\end{align}
where we have used the abbreviations
\begin{align}
 A&=W''+r_1-\frac{1}{2}(Y_1'+Y_2)q^2\notag\\
   B&=Z'^2+r_2 Z_k'^2(\bar{\phi})\notag\\
   C&=B q^2-\frac{1}{4} F^3 Y_2'''+F^2 \left(Z''^2+Z'''
   Z'\right)+\frac{1}{2}F \left(2W'''- q^2 \left(2
   Y_2'+Y_1''\right)\right).
\label{abbrev3}
\end{align}

\section{(Pseudo-)spectral methods}
\label{sec:appendix3}

We obtained the numerical results in this work in part with so-called (pseudo-)spectral methods. Spectral methods were studied from a mathematical point of view
already some decades ago, however, they were only applied in certain fields of physics up to now, e.g. in numerical relativity, meteorology or fluid mechanics.
The basic idea behind spectral methods is to expand the solution into orthogonal polynomials which should be chosen to fit the problem.
A well-known example is the Fourier series of a periodic function. In our case, we chose a combination of Chebyshev and rational Chebyshev polynomials in order
to resolve the operators globally. On the other hand, in RG-time-direction, we chose to map the (infinite) time axis onto a finite interval,
then slicing it into smaller pieces and apply a Chebyshev spectralization in this direction. With a stabilized Newton-Raphson iteration scheme,
we demanded that the flow equations are satisfied on collocation points up to a certain tolerance. This twofold application of spectral methods was considered
too expensive in former times, but thanks to the progress in computing power, it is feasible now. This point is also undermined
by the recent application of this method to gain exact solutions to the Einstein field equations for axisymmetric and stationary space times \cite{Ansorg:2003br,Macedo:2014bfa}.

The reason to use spectral methods is their extraordinary speed of convergence. For well-behaved functions, a spectral method may convergence exponentially,
i.e. faster than any power law. Another advantage is that the expansion coefficients give a rough estimate of the maximal error in the interpolation of the solution.
A general rule of thumb is that the error is bounded by roughly the absolute value of the last coefficient retained. For an extensive review of spectral methods, 
see e.g. \cite{Boyd:ChebyFourier}.
 
\bibliographystyle{unsrt}   
\bibliography{References}

\end{document}